\documentclass[fleqn,10pt, tikz,border=10pt]{wlscirep}
\usepackage{amsmath}
\usepackage{graphicx}  
\usepackage{nameref}
\usepackage{hyperref}
\usepackage{setspace}
\usepackage{lineno}
\usepackage[ruled,linesnumbered]{algorithm2e}
\SetAlgoNoEnd 
\usepackage{bibspacing}

\usepackage{caption}

\usepackage{subcaption}  
\usepackage[utf8]{inputenc}
\usepackage{tabularx,longtable,multirow,makecell}
\usepackage[figurename=Fig.]{caption}
\usepackage[nolist]{acronym}

\usepackage{multibib}
\newcommand{\edited}[1]{#1} 

\newcites{methods}{Methods References}

\newif\ifinsupp
\insuppfalse
\providecommand{\noopsort}[1]{}

\title{%
\begin{minipage}[c]{1.7cm}%
\includegraphics[width=1.7cm]{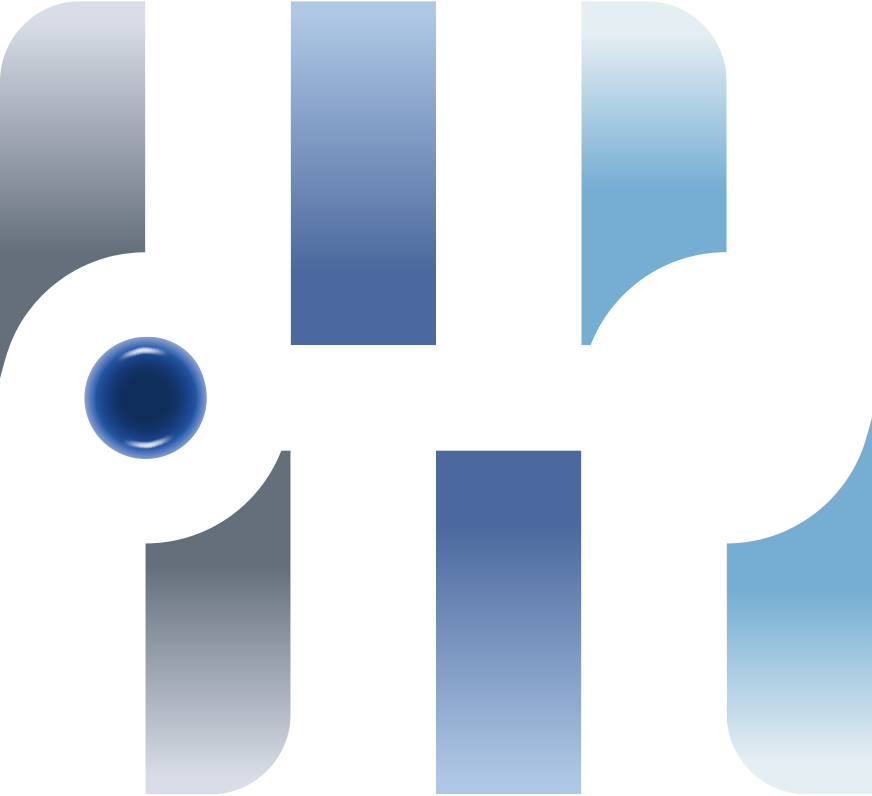}%
\end{minipage}%
\hspace{0.4em}%
\begin{minipage}[c]{\dimexpr\linewidth-2.0cm\relax}%
\raggedright \noindent CaliPPer: quantifying, predicting and improving AI model performance for binding prediction%
\end{minipage}%
}

\author[1,*,$\dag$]{Jian-Qing Zheng}
\author[2,*]{Hantao Lou}
\author[1,*]{Zinan Yin}
\author[1]{Sam Farrar}
\author[1,3]{Yuze Zhou}
\author[1,3]{Elie Antoun}
\author[4]{Xiangxi Wang}
\author[1,2,5,$\dag$]{Xuetao Cao}
\author[1,3,$\dag$]{Tao Dong}

\affil[1]{Chinese Academy of Medical Science (CAMS) Oxford Institute (COI), University of Oxford, Oxford, UK}
\affil[2]{State Key Laboratory of Medicinal Chemical Biology, Institute of Immunology, College of Life Sciences, Nankai University, Tianjin, CN}
\affil[3]{Center for Translational Immunology, Nuffield Department of Medicine, University of Oxford, UK}
\affil[4]{Key Laboratory of Infection and Immunity, National Laboratory of Macromolecules, Institute of Biophysics, Chinese Academy of Sciences, Beijing, CN}
\affil[5]{Department of Immunology, Center for Immunotherapy, Peking Union Medical College, Chinese Academy of Medical Sciences, Beijing, CN}
\affil[*]{Equal contribution}
\affil[$\dag$]{Corresponding author: jianqing.zheng@ndm.ox.ac.uk; caoxt@immunol.org; tao.dong@ndm.ox.ac.uk}

\begin{abstract}
\edited{Binding prediction models accelerate therapeutic antibody and TCR discovery, but their performance on new datasets is unpredictable, often leading to low discovery rates. Density-ratio methods (PAPE, M-CBPE) provide label-free performance estimation for binary classification, but their assumptions and aggregate-only outputs limit binding prediction on neoepitopes, antigen variants and chemical scaffolds. Here we present CaliPPer (Calibration and Prediction of Performance), a post-hoc framework pairing a multi-chain Sample-to-Domain Distance (S2DD) with distance-aware Bayesian recalibration, operating at three resolutions: generalisability score, aggregate performance prediction, and per-sample confidence. Across ten models, eight architectures and two immune-receptor domains, CaliPPer attains distance--performance correlations $|r|=0.80\text{--}0.92$, predicts AUROC/AP/F1 with mean absolute errors $0.008\text{--}0.070$, and improves AUROC by up to $+0.20$ on unseen epitopes/variants. Applied retrospectively to five published TCR, BCR, MHC--peptide and small-molecule studies, CaliPPer raises true discovery rates in all five (e.g.\ $0/5 \to 3/5$ confirmed neoantigens), providing a triage layer between computational prediction and experimental validation.}
\end{abstract}

\begin{document}

\flushbottom
\maketitle

\thispagestyle{empty}

\acrodef{TCR}[TCR]{T-cell receptor}
\acrodef{BCR}[BCR]{B-cell receptor}
\acrodef{S2DD}[S2DD]{Sample-to-Domain Distance}
\acrodef{CaliPPer}[CaliPPer]{Calibration and Prediction of Performance}
\acrodef{AP}[AP]{average precision}
\acrodef{MCC}[MCC]{Matthews correlation coefficient}
\acrodef{AUROC}[AUROC]{area under the ROC curve}
\acrodef{RBD}[RBD]{receptor binding domain}
\acrodef{MAE}[MAE]{mean absolute error}
\acrodefplural{MAE}[MAEs]{mean absolute errors}
\acrodef{PAPE}[PAPE]{Probabilistic Adaptive Performance Estimation}
\acrodef{MCBPE}[M-CBPE]{Multi-Calibrated Confidence-Based Performance Estimation}
\acrodef{PPV}[PPV]{positive predictive value}
\acrodef{NPV}[NPV]{negative predictive value}
\acrodef{TDR}[TDR]{true discovery rate}
\acused{AP}
\acused{AUROC}


\section*{Main}


\edited{Anticipating the failure modes of artificial intelligence models on out-of-distribution data is increasingly recognised as a central challenge across scientific disciplines~\cite{zhou2026general}, from drug activity predictors that lose precision on novel chemical structures~\cite{wong2024antibiotics} to protein structure prediction models encountering highly flexible loops. In computational biology, unreliable model predictions have practical consequences: each failed binding validation consumes reagents, occupies bench time, and delays therapeutic development. Immune receptor binding prediction is a particularly acute case. T cell receptor (\ac{TCR}) and B cell receptor (\ac{BCR}) models, trained on known receptor--antigen pairs, achieve high precision on benchmark datasets~\cite{nettcr,atmtcr,xbcrnet}, but this performance does not transfer to novel targets. Systematic benchmarks have repeatedly shown that \ac{TCR}--epitope binding predictors fail on epitopes absent from training~\cite{grazioli2022attentive,tcrpmhcbenchmark2024}; 23 of 50 models achieved AUPRC (also known as average precision, AP) $\leq 0.5$ on held-out epitopes~\cite{lu2025assessment}. The same generalisation gap affects \ac{BCR} models across antigen variants and pathogen classes~\cite{zhang2022deepaai,wang2024rleaai}.}

\edited{The challenge is not that models perform poorly in general, but that we cannot predict \emph{how poorly they will perform on a specific new dataset}. This creates three unmet needs at progressively finer resolution. Model developers need a metric to quantify how generalisable their model is, to guide optimisation and compare architectures. Researchers applying models need a prediction of how the model will perform on their specific new cohort, before running experiments. And screening teams need per-prediction confidence to prioritise which candidates to test experimentally.}

\edited{Currently, models are compared by aggregate accuracy on held-out data that shares the training distribution; predictions on new datasets can only be validated by wet-lab experiments; and model-native probability scores are poorly calibrated on out-of-distribution data~\cite{guo2017calibration,ovadia2019trust}. The benchmark that exposed the generalisability crisis found no correlation between average edit distance and performance ($|r| < 0.1$)~\cite{lu2025assessment}, indicating that na\"ive distance metrics cannot capture the relationship between distributional shift and performance. A recent \ac{TCR} antigen-discovery study did report model performance anticorrelating with a peptide-recognition-profile distance between held-out and training receptors ($\rho \approx -0.78$)~\cite{wang2026prptcr}; this correlation describes a cross-dataset population trend rather than a per-dataset estimate, leaving open how distance can be converted into a per-dataset performance prediction for a specific new cohort.}

\edited{Two biological features of binding prediction explain why label-free performance estimation has remained difficult in this domain. First, novel epitopes and antigen variants occupy regions of sequence space the training set never visited. Second, the rules that govern binding can differ for a new epitope; the model's input-to-output mapping itself shifts between training and test. Existing label-free estimators developed for general binary classification, \ac{PAPE}~\cite{bialek2024pape} and \ac{MCBPE}~\cite{bialek2024model}, address neither feature: they re-weight calibration predictions under two assumptions, that training and test data cover the same sequence space (\emph{support overlap}) and that the binding rules stay constant (the \emph{covariate-shift-only} assumption). Novel epitopes and antigens violate both (Supplementary Note~9). On binding problems, these gaps cause density-ratio predictions to collapse toward a single average value across all test sets, diverging most sharply from truth on the novel targets that matter most for therapeutic development. These methods also return only an aggregate accuracy estimate, with no per-sample confidence to support candidate prioritisation.}

\begin{figure}[h]
    \centering
    \includegraphics[width=0.9\linewidth]{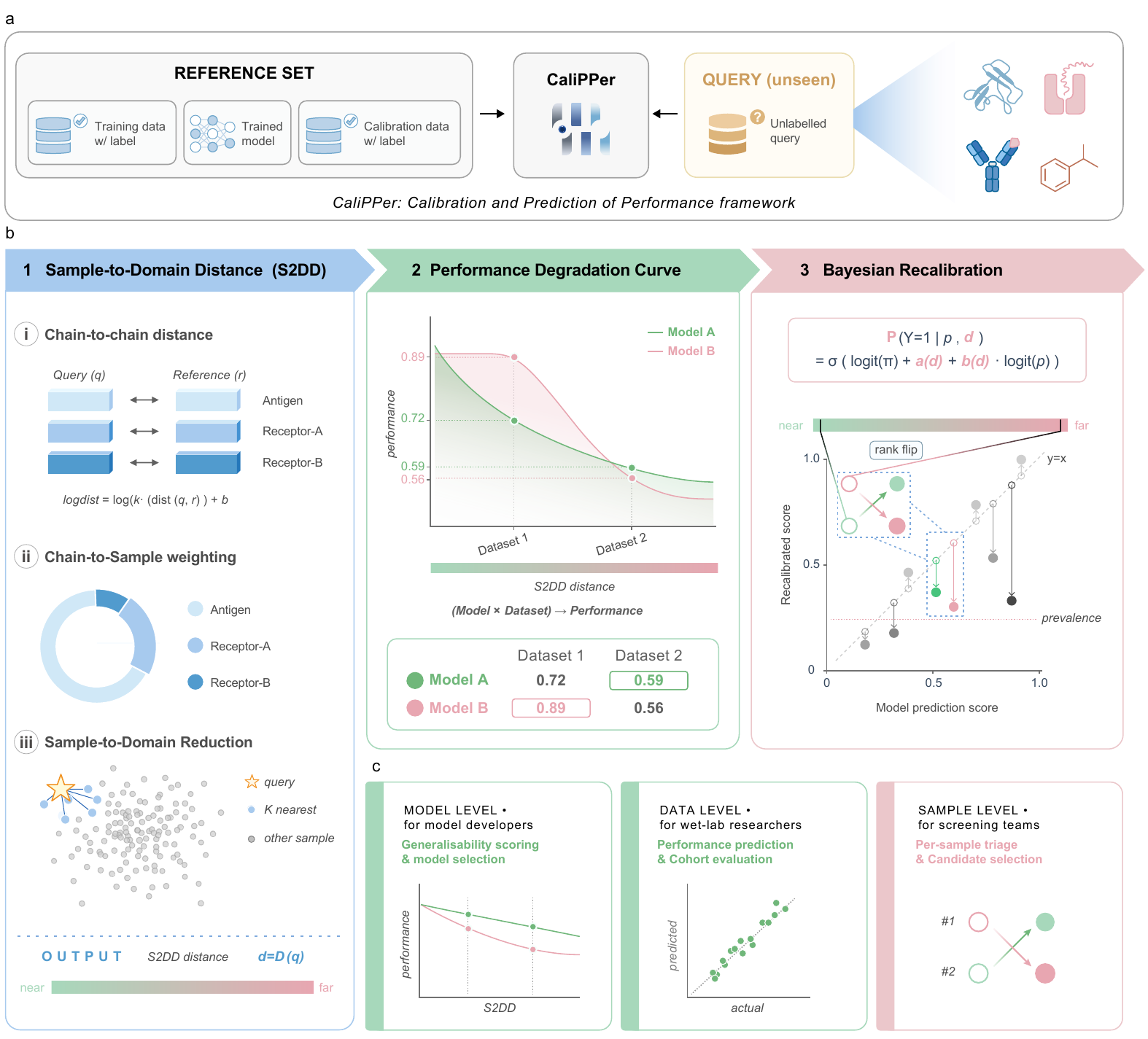}
    \caption{\edited{\textbf{CaliPPer framework overview.} Binding-prediction classifiers (\ac{TCR}--epitope, \ac{BCR}--antigen, MHC--peptide and drug--target) can degrade unpredictably on novel test data in ways that general-domain label-free performance estimators (PAPE, M-CBPE) cannot anticipate when applied to binding, owing to two structural assumptions (support overlap, covariate shift) that break down on novel epitopes/variants/scaffolds, together with an aggregate-only output limitation that precludes per-sample triage. CaliPPer (Calibration and Prediction of Performance) is a post-hoc framework that closes these gaps by integrating a novel multi-chain Sample-to-Domain Distance (\ac{S2DD}) with distance-conditioned aggregate prediction and per-sample Bayesian recalibration. \textbf{a}, Workflow. CaliPPer takes a reference set (training data, trained model, labelled calibration set) and unlabelled query samples (\ac{TCR}, \ac{BCR}, MHC--peptide or small-molecule) and produces distance-conditioned outputs at three resolutions. \textbf{b}, Three CaliPPer components. (1) \emph{\ac{S2DD} distance:} (i) chain-to-chain LogDist ($\mathrm{logdist} = \log(k\!\cdot\!\mathrm{dist}(q,r) + b)$) per chain between query and reference; (ii) $\sigma \!\cdot\! C$ concentration weighting that upweights whichever chain has the most concentrated training distribution; (iii) top-$K$ nearest-neighbour reduction to a scalar per query. Smaller \ac{S2DD} denotes samples nearer to training; the base distance is modular (Levenshtein, BLOSUM, ESM-2, RMSD or Morgan fingerprint), spanning sequence- and structural-level signal. (2) \emph{Performance Degradation Curve:} per-bin performance fitted as a function of \ac{S2DD} across calibration datasets, yielding a (Model~$\times$~Dataset)~$\to$~Performance map that characterises how each model's reliability decays as a function of distance. (3) \emph{Bayesian Recalibration:} per-sample recalibrated scores derived analytically from distance-dependent \ac{PPV} and \ac{NPV} via $P(Y{=}1 \mid p, d) = \sigma(\operatorname{logit}(\pi) + a(d) + b(d)\!\cdot\!\operatorname{logit}(p))$; distance-dependent $a(d), b(d)$ break the rank-preserving constraint of standard Platt scaling, permitting \ac{AUROC} and \ac{AP}, not only ECE, to improve. \textbf{c}, Three application levels. \emph{Model level} (for model developers): distance-resolved generalisability profiles for architecture comparison and selection. \emph{Data level} (for wet-lab researchers): aggregate performance prediction on unlabelled query cohorts without test labels, supporting cohort-prioritisation decisions before validation. \emph{Sample level} (for screening teams): Bayesian-recalibrated per-sample confidence scores for candidate triage; in retrospective application to five published studies (\ac{TCR} neoantigen, \ac{BCR} antigen, MHC--peptide and small-molecule), this raised top-$k$ true discovery rates in all five (e.g.\ $0/5 \to 3/5$ confirmed immunogenic neoantigens in the deepAntigen pipeline) without retraining the underlying model.}}
    \label{fig:overview}
\end{figure}

\edited{Here we present \ac{CaliPPer}, a model-agnostic framework that addresses these gaps by anchoring performance estimation on a binding-specific distance signal. \ac{CaliPPer} characterises performance as a function of \ac{S2DD}, a distributional distance metric that quantifies how far each test sample lies from the training distribution directly in sequence space (Fig.~\ref{fig:overview}); fitting a mathematical curve to this distance--performance relationship captures the biology that density-ratio methods alone cannot. Building on the principle that sequence similarity carries functional information in immune receptors~\cite{dash2017quantifiable}, S2DD combines receptor and antigen chain distances using a weighting scheme ($\sigma \!\cdot\! C$) that emphasises chains with concentrated training distributions (high $C$) and informative inter-sequence distance variation (high $\sigma$); small differences on these chains then reliably signal novelty. The base distance is modular, supporting Levenshtein edit distance~\cite{levenshtein1966binary}, BLOSUM amino-acid substitution scores~\cite{henikoff1992amino}, ESM-2 protein language model embeddings~\cite{lin2023esm2}, or structural (RMSD) formulations; the combining step is the same regardless of which base distance is used.}

\edited{\ac{CaliPPer} is, to our knowledge, the first framework that converts the distance--performance relationship in immune-receptor binding into a quantitative performance estimate at three operating resolutions, addressing a gap that aggregate-only label-free estimators leave open in this domain. At the \emph{model level}, the full degradation curve characterises each model's reliability across the distance spectrum; comparing curves between models reveals which architecture performs the best in each distance regime, enabling distance-aware architecture selection when the distributional shift of a new dataset can be estimated. At the \emph{data level}, \ac{CaliPPer} predicts aggregate performance on unlabelled datasets by fitting a distance-conditioned degradation curve to the \ac{S2DD}–performance relationship and combining it with a \ac{PAPE}-derived density-ratio correction. The \ac{S2DD} curve provides a reliable estimate even when test sequences fall in sequence regions not covered by training, or when the binding rules differ for new targets; the density-ratio term refines the prediction in the regime where neither of these conditions holds. At the \emph{sample level}, \ac{CaliPPer} applies diagnostic-test logic: just as a clinical assay's \ac{PPV} and \ac{NPV} depend on disease prevalence, a binding model's predictive value depends on each test sample's distance from training. CaliPPer fits distance-dependent PPV($d$) and NPV($d$) curves on a labelled calibration set and uses them to adjust each prediction's confidence. Each prediction's adjustment combines distance-dependent PPV($d$) and NPV($d$) through a closed-form Bayesian update: well-supported predictions retain their scores; predictions at distances where both confidences weaken are pulled toward the prior prevalence; and at distances where the local model becomes anti-informative (Supplementary Note~7), the update crosses the prior, mapping raw positives below it and raw negatives above it. Because the adjustment varies with distance rather than being applied uniformly, CaliPPer re-ranks candidates by distance-conditioned reliability rather than by raw score alone (Supplementary Note~7). Together, the three resolutions convert a model's prediction on novel data from an unverifiable claim into a quantitative decision.}

\edited{We validate \ac{CaliPPer} across ten prediction models from eight architectural families (CNN~\cite{nettcr,xbcrnet}, LSTM~\cite{springer2020prediction,wang2024rleaai}, GCN~\cite{zhang2022deepaai}, attention/Transformer~\cite{atmtcr,wu2024tcrbert,ullanat2026learning}, state-space model~\cite{liu2024mambaaai}, SVM~\cite{wu2024tcrbert}, random forest~\cite{pham2023epitcr}, MLP~\cite{ullanat2026learning}; Supplementary~Table~S1). Validation spans two domains: \ac{TCR}--epitope binding (five models, 40{,}516 samples, 783 epitopes) and \ac{BCR}--antigen binding (five models, SARS-CoV-2 and influenza, 80 antigen variants). \ac{CaliPPer} predicts held-out performance with \acp{MAE} of 0.008--0.070 across cross-validation (CV) and cross-test (CT, independent test sets reserved entirely from training) on both \ac{TCR} and \ac{BCR}, and improves discrimination by up to +0.20 \ac{AUROC} without model retraining. Retrospective analysis of five independent published studies (spanning \ac{TCR} immunogenicity~\cite{que2025deepantigen}, \ac{BCR} antigen specificity~\cite{xbcrnet}, pan-allele \ac{TCR} meta-learning~\cite{gao2023panpep}, MHC presentation~\cite{albert2024bigmhc} and small-molecule antibiotic activity~\cite{wong2024antibiotics}) shows the distance--performance relationship transfers across receptor classes, prediction tasks and to molecular-fingerprint representations, improving true discovery rates in all five studies (e.g.\ 0/5 $\to$ 3/5 for neoantigen candidates, 22/50 $\to$ 29/50 for MHC immunogenicity) without additional training, datasets, or experiments.}

\section*{Results}

\subsection*{S2DD reveals consistent performance degradation across architectures and domains}

\begin{figure}[t]
    \centering
    \includegraphics[width=\linewidth]{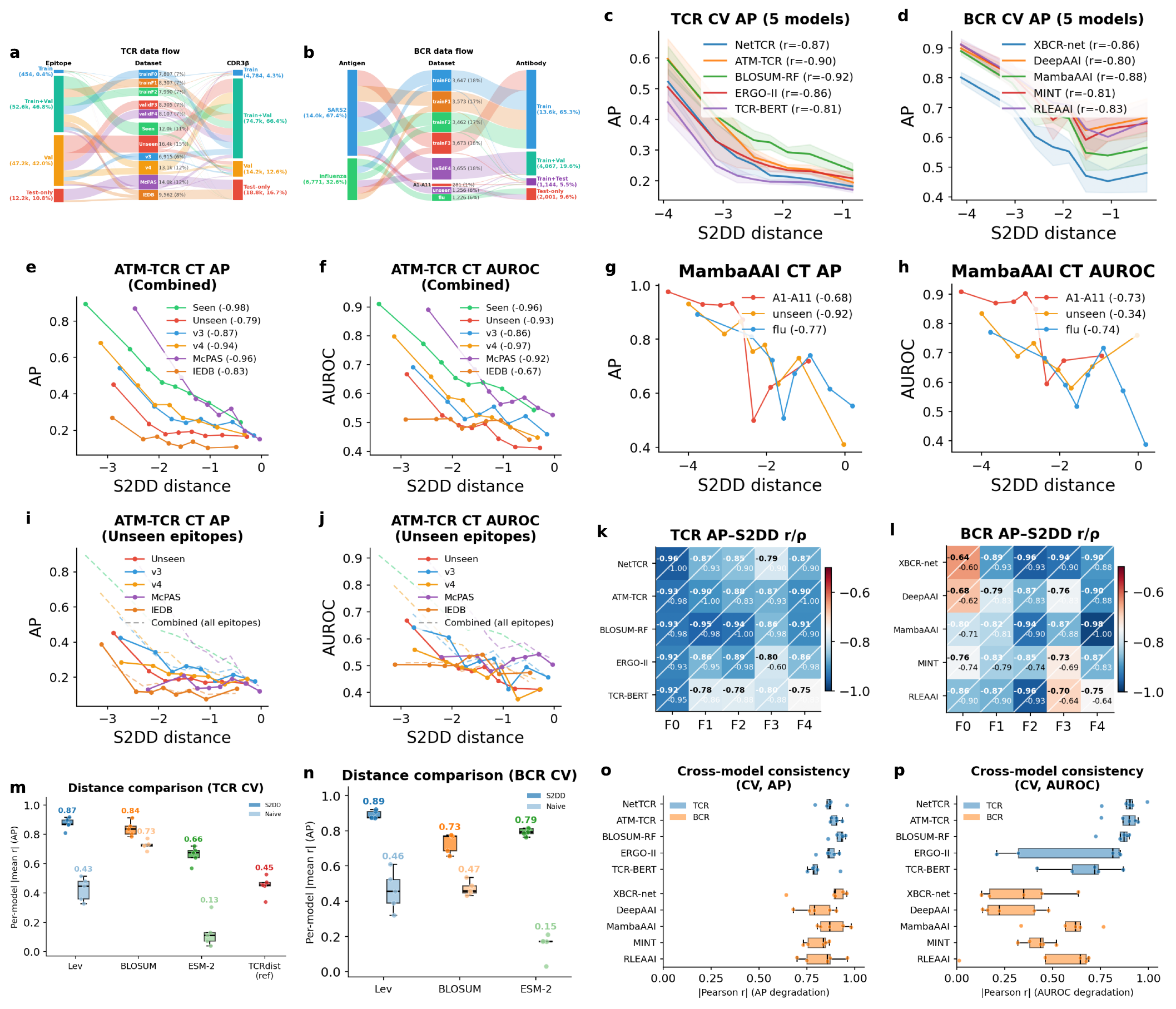}
    \caption{\edited{\textbf{Performance degradation with distributional shift is systematic across 10 models and 2 receptor types.} \textbf{a, b}, TCR and BCR data-flow Sankey diagrams (epitope/antigen partition across training, validation and test-only sets). \textbf{c}, TCR cross-validation AP degradation for 5 models (mean $\pm$ s.d.\ over 5 folds); all decline with S2DD distance. \textbf{d}, As \textbf{c}, 5 BCR models. \textbf{e, f}, ATM-TCR cross-test AP and AUROC across 6 test sets (one line per set). \textbf{g, h}, As \textbf{e, f} for MambaAAI (BCR, 3 test sets). \textbf{i, j}, ATM-TCR cross-test on unseen epitopes (solid) overlaid on the combined-data curves of \textbf{e, f} (dashed); the agreement shows degradation is consistent between seen and unseen epitopes. \textbf{k}, TCR CV AP correlation heatmap (5 models $\times$ 5 folds; upper triangle Pearson $r$, lower Spearman $\rho$); all cells negative, most $|r|,|\rho|>0.7$. \textbf{l}, As \textbf{k} for BCR. \textbf{m}, Distance-metric comparison (TCR CV): per-model $|$mean Pearson $r|$ for AP degradation, S2DD multi-chain versus a naive single-chain CDR3$\beta$ baseline, as paired boxplots for Levenshtein, BLOSUM and ESM-2; TCRdist (Dash et al., CDR3$\alpha$+CDR3$\beta$) is shown as a single reference box. The naive baseline is the mean single-chain distance over all training, with no S2DD aggregation (no log, z-normalisation, top-$K$ or nearest-neighbour selection; BLOSUM additionally length-orthogonalised; Supplementary~Note~6). S2DD outperforms the single-chain baseline for every metric. \textbf{n}, As \textbf{m} for BCR CV, 5 models (heavy-chain naive). \textbf{o, p}, Per-model degradation strength: boxplots of $|$Pearson $r|$ across folds for AP and AUROC, all 10 models.}}
    \label{fig:degradation}
\end{figure}

\edited{To establish whether prediction performance degrades systematically with distributional shift, we evaluated five \ac{TCR}--epitope and five \ac{BCR}--antigen binding models under matched train, validation and held-out cross-test partitions. For \ac{TCR}, 783 epitopes from diverse pathogens were split into seen-epitope and held-out unseen-epitope sets, supplemented by the external McPAS-TCR and IEDB-SARS resources and the in-house v3 and v4 sets (Fig.~\ref{fig:degradation}a). For \ac{BCR}, 80 SARS-CoV-2 and influenza antigen variants were split by antibody across SARS-CoV-2 (A1--A11), unseen Omicron-era variants, and influenza (Fig.~\ref{fig:degradation}b). Across all ten models, performance declined systematically with S2DD distance, with negative Pearson correlations in every model tested.}

\edited{In five-fold cross-validation, the five \ac{TCR} models all showed negative S2DD--\ac{AP} correlations, with $|r|$ between 0.81 and 0.92 (Fig.~\ref{fig:degradation}c): NetTCR~\cite{nettcr} (CNN), ATM-TCR~\cite{atmtcr} (attention), BLOSUM-RF~\cite{pham2023epitcr} (Random Forest), ERGO-II~\cite{springer2020prediction} (LSTM) and TCR-BERT~\cite{wu2024tcrbert} (Transformer+SVM). The five \ac{BCR} models degraded equivalently across 80 antigens under antibody-stratified cross-validation, with $|r|$ between 0.80 and 0.88 (Fig.~\ref{fig:degradation}d): XBCR-net~\cite{xbcrnet} (CNN), DeepAAI~\cite{zhang2022deepaai} (GCN), MambaAAI~\cite{liu2024mambaaai} (state-space model), MINT~\cite{ullanat2026learning} (Transformer+MLP) and RLEAAI~\cite{wang2024rleaai} (LSTM).}

\begin{figure}[t]
    \centering
    \includegraphics[width=\linewidth]{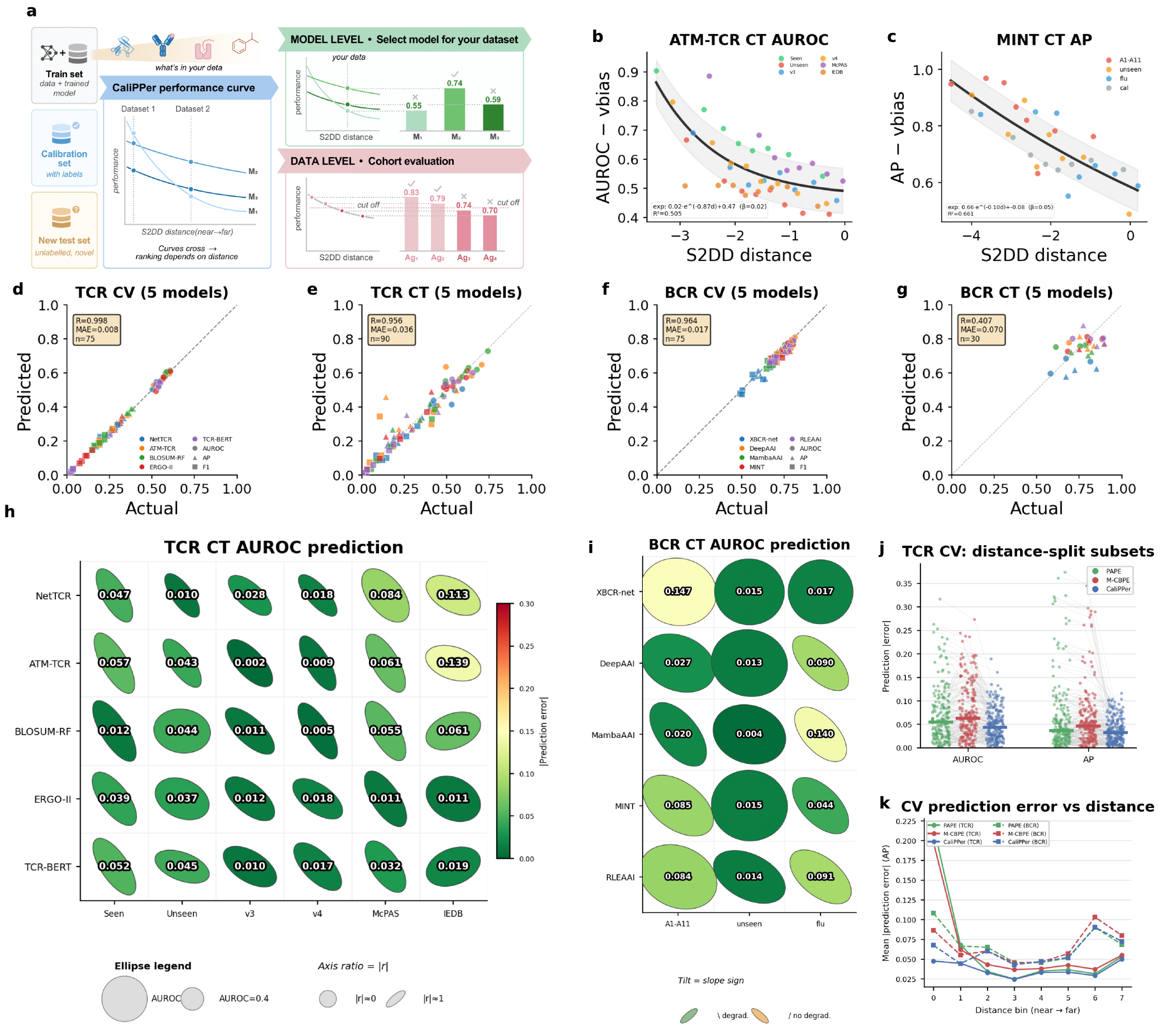}
    \caption{\edited{\textbf{CaliPPer accurately predicts model performance on new datasets across 10 models.} \textbf{a}, Conceptual overview of CaliPPer's performance prediction supporting two applications: model-level architecture selection (top right) and data-level cohort triage (bottom right). \textbf{b}, Representative distance--performance curve for a TCR model (ATM-TCR, cross-test AUROC); the bias-corrected decay curve is fitted on distance-binned calibration data, and each point is one held-out cross-test set (dataset-level, coloured by test set). \textbf{c}, Same for a BCR model (MINT, cross-test AP). \textbf{d}, Predicted vs actual performance for TCR cross-validation (5 models $\times$ 5 folds $\times$ 3 metrics (AUROC, AP, F1) $=$ 75 points). \textbf{e}, Predicted vs actual for TCR cross-test (5 models $\times$ 6 test sets $\times$ 3 metrics $=$ 90 points, with each test set held out in turn). \textbf{f}, Same as \textbf{d} for BCR cross-validation. \textbf{g}, Same as \textbf{e} for BCR cross-test (5 models $\times$ 2 metrics (AUROC, AP) $\times$ 3 test sets $=$ 30 points, calibrated on BCR cross-validation fold~4). \textbf{h}, Per-model per-test-set prediction-error map for TCR cross-test (5 models $\times$ 6 test sets, 30 cells). Each ellipse is one model--test-set combination: area is proportional to the actual AUROC (larger $=$ better-performing model); colour and the inner number give the absolute prediction error $|$predicted $-$ actual$|$ (green low to red high); aspect ratio encodes the strength of the distance--performance degradation correlation ($|r|$; more elongated $=$ stronger); and long-axis orientation encodes the degradation-slope sign (down-tilted for negative slope, up-tilted for positive). \textbf{i}, Same as \textbf{h} for BCR cross-test (5 models $\times$ 3 test sets, 15 cells). \textbf{j}, Prediction $|$error$|$ decomposition: the density-ratio base estimate alone (PAPE; M-CBPE variant for reference) versus the full CaliPPer (density-ratio base + S2DD-anchored correction), on TCR cross-validation with distance-split subsets (S2DD-distance bins, not epitope): per-subset prediction $|$error$|$ grouped by metric (AUROC, AP); points are individual subsets; faint lines pair same-subset across methods; horizontal bars mark each method's median. The full CaliPPer achieves the lowest median error and fewer high-error outliers. \textbf{k}, Mean prediction $|$error$|$ (AP) per distance bin (near $\to$ far from training) for the three methods (CaliPPer, PAPE, M-CBPE), shown for TCR (solid lines) and BCR (dashed lines).}}
    \label{fig:prediction}
\end{figure}

\edited{The same pattern held on held-out cross-test data. Across six independent \ac{TCR} test sets, including the external McPAS-TCR~\cite{mcpas2017} and IEDB-SARS~\cite{vita2019iedb} datasets, both \ac{AP} and \ac{AUROC} declined with increasing S2DD distance (Fig.~\ref{fig:degradation}e,f). Restricting to unseen epitopes, where distributional shift is most severe, gave correlations of comparable magnitude, and the unseen-epitope curves closely tracked the combined-data curves from the same test sets (Fig.~\ref{fig:degradation}i,j, solid versus dashed). This consistency between seen and unseen epitopes is the empirical foundation for CaliPPer's prediction and recalibration: curves fitted on calibration data that include seen epitopes transfer to test data dominated by unseen epitopes. \ac{BCR} cross-test data showed equivalent behaviour across SARS-CoV-2, unseen variants and influenza (Fig.~\ref{fig:degradation}g,h). Aggregated across all model--fold combinations, \ac{AP} correlations were negative in every cell (Fig.~\ref{fig:degradation}k,l), and all ten models achieved $|r| > 0.6$ on \ac{AP} (Fig.~\ref{fig:degradation}o). \ac{BCR} \ac{AUROC} correlations were weaker (mean $|r| = 0.28$--$0.58$; Fig.~\ref{fig:degradation}p) because \ac{BCR} \ac{AUROC} varies over a compressed range across folds (s.d.\ $= 0.013$), roughly half of the other three domain--metric combinations (\ac{BCR} \ac{AP} $0.016$, \ac{TCR} \ac{AUROC} $0.022$, \ac{TCR} \ac{AP} $0.019$; all $\approx 0.02$), leaving insufficient variation for distance to correlate with; this compressed-range effect is also visible in the tight \ac{BCR} cross-test prediction scatter (Fig.~\ref{fig:prediction}g).}

\edited{S2DD's combination of distances across all three chains (\ac{TCR}: peptide, CDR3$\alpha$, CDR3$\beta$; \ac{BCR}: heavy, light, antigen variant) was essential to this consistency. Naive single-chain baselines (the simple mean edit distance to all training sequences from a single chain: CDR3$\beta$ for \ac{TCR}, heavy chain for \ac{BCR}, with no chain-weighting or nearest-neighbour selection) showed substantially weaker correlations. For example, on \ac{TCR} Levenshtein distance, mean $|r|$ was 0.87 with S2DD versus 0.43 with the single-chain CDR3$\beta$ baseline (Fig.~\ref{fig:degradation}m,n). The multi-chain advantage held across every base distance metric with a single-chain comparator (Levenshtein, BLOSUM and ESM-2; per-test-set statistics in Supplementary~Table~S15), confirming that S2DD's chain-weighting, rather than the choice of base distance, drives the result. For reference, the genuine TCRdist receptor distance (Dash et al.; CDR3$\alpha$+CDR3$\beta$) is shown as an additional single-chain-style baseline (Fig.~\ref{fig:degradation}m).}

\subsection*{CaliPPer predicts dataset-level performance on unlabelled data}

\edited{To test whether this distance--performance relationship can predict aggregate model performance on unlabelled test data, we applied \ac{CaliPPer} to new test sets (Fig.~\ref{fig:prediction}a). Performance decayed approximately exponentially with S2DD distance from training; \ac{CaliPPer} fits this decay using exponential curves on labelled calibration data with an additional term that accounts for variation in average predicted probability across calibration sets. \ac{CaliPPer} then predicts performance on new test sets from their S2DD distance profile alone (Fig.~\ref{fig:prediction}b,c show representative \ac{TCR} and \ac{BCR} fits; Methods).}

\edited{Predictions tracked observed performance closely across all four evaluation settings: \ac{TCR} cross-validation ($R = 0.998$, \ac{MAE} $= 0.008$, $n = 75$; Fig.~\ref{fig:prediction}d); \ac{TCR} cross-test ($R = 0.956$, \ac{MAE} $= 0.036$, $n = 90$; Fig.~\ref{fig:prediction}e); \ac{BCR} cross-validation ($R = 0.964$, \ac{MAE} $= 0.017$, $n = 75$; Fig.~\ref{fig:prediction}f); and \ac{BCR} cross-test ($R = 0.407$, \ac{MAE} $= 0.070$, $n = 30$; Fig.~\ref{fig:prediction}g). The lower \ac{BCR} cross-test $R$ reflects the narrow \ac{AUROC} range spanned by the three \ac{BCR} test sets (s.d.\ $= 0.09$ versus $0.20$ for \ac{TCR}), which mathematically limits the achievable correlation. Absolute prediction error nonetheless remained low (\ac{MAE} $= 0.070$, within 7 percentage points of observed performance). The calibration domain itself (\ac{BCR} cross-validation) was not range-compressed ($R = 0.964$, \ac{MAE} $= 0.017$). Per-model per-test-set heatmaps confirmed that most cells had absolute prediction error below 0.05 (Fig.~\ref{fig:prediction}h,i).}

\edited{CaliPPer's binding-specific design hinges on closing two structural gaps that the density-ratio paradigm leaves open on novel epitopes and variants. To test whether the distance-conditioned correction closes these support-overlap and covariate-shift gaps (Supplementary~Note~9), we compared the full \ac{CaliPPer} to its density-ratio component evaluated alone (the \ac{PAPE}~\cite{bialek2024pape} formulation) and to the \ac{MCBPE}~\cite{bialek2024model} density-ratio variant. The full \ac{CaliPPer} predicted with lower errors and fewer high-error outliers across distance-split subsets (Fig.~\ref{fig:prediction}j). \ac{PAPE} alone produced \ac{AUROC} predictions that compressed to a $0.13$ span around the calibration mean versus the $0.37$ span of actual values (Fig.~\ref{fig:prediction}k; Supplementary~Table~S18); this range compression is the empirical signature of the two structural gaps. CaliPPer's distance-conditioned correction closes this gap, with prediction error at the nearest-distance bin $4.9\,\times$ larger without the distance correction than with it on \ac{AUROC}, and $4.6\,\times$ larger on \ac{AP}. At far distances, errors are small for all methods because both the calibration mean and the model's true performance converge toward chance, so the metric itself is least discriminating in this regime; the near-distance regime, where models retain signal, is the more informative comparison.}

\subsection*{CaliPPer predicts per-epitope and per-variant performance}

\begin{figure}[t]
    \centering
    \includegraphics[width=\linewidth]{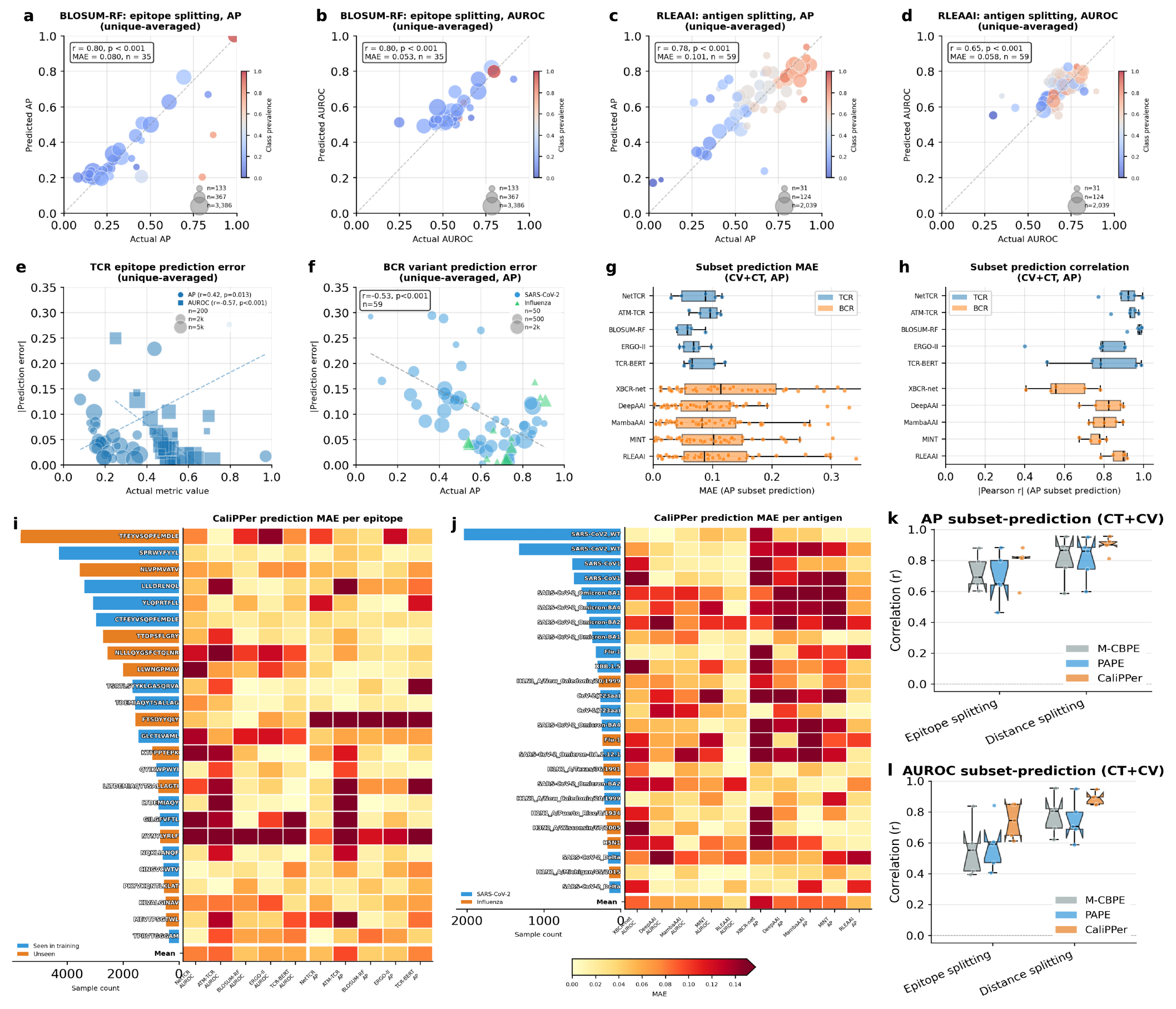}
    \caption{\edited{\textbf{CaliPPer predicts per-epitope and per-variant performance across 10 models.} \textbf{a}, Predicted versus actual AP for individual TCR epitopes (BLOSUM-RF model; metric computed by per-epitope averaging). Bubble size proportional to sample count; colour encodes class prevalence. \textbf{b}, Same as \textbf{a} for AUROC. \textbf{c, d}, Same as \textbf{a, b} for BCR antigen variants (RLEAAI model). \textbf{e}, Model-averaged prediction $|$error$|$ versus actual performance for TCR epitopes, shown for AP (positive correlation, $r = 0.39$) and AUROC (negative correlation, $r = -0.53$). The negative AUROC correlation indicates higher prediction error on epitopes where the underlying model achieves lower AUROC; the positive AP correlation reflects ceiling effects on high-AP epitopes. \textbf{f}, Model-averaged AP prediction $|$error$|$ versus actual AP for BCR variants, coloured by pathogen (SARS-CoV-2 versus influenza). Negative correlation ($r = -0.55$) indicates higher prediction error on variants where the underlying model achieves lower AP. \textbf{g}, Subset prediction MAE distribution across 10 models (5 TCR blue, 5 BCR orange), combining cross-validation and cross-test evaluations. \textbf{h}, Subset prediction correlation ($|r|$) distribution by subset type (epitope splitting versus distance splitting) and method (CaliPPer's S2DD-based prediction, PAPE, M-CBPE). CaliPPer shows the highest median correlation and tightest spread, particularly on distance-split subsets. \textbf{i}, Per-epitope prediction error heatmap for TCR (25 epitopes $\times$ 10 columns: 5 models showing AUROC prediction error and 5 models showing AP prediction error). A side bar chart indicates whether each epitope was seen (blue) or unseen (orange) in training. \textbf{j}, Same as \textbf{i} for BCR antigen variants (20 variants $\times$ 10 columns: 5 models $\times$ 2 metrics). \textbf{k}, AP prediction correlation by subset type (epitope splitting versus distance splitting) for the three methods (CaliPPer, PAPE, M-CBPE). \textbf{l}, Same as \textbf{k} for AUROC.}}
    \label{fig:subset_prediction}
\end{figure}

\edited{To test whether the same distance--performance relationship predicts performance at finer resolution, we evaluated \ac{CaliPPer} on individual epitopes and antigen variants. For \ac{TCR}, per-epitope predictions correlated with observed values at $r = 0.78$ for \ac{AP} and $r = 0.71$ for \ac{AUROC} (\ac{MAE} $= 0.088$, $n = 35$ epitopes; Fig.~\ref{fig:subset_prediction}a,b); for \ac{BCR}, per-variant predictions achieved $r = 0.77$ for \ac{AP} and $r = 0.65$ for \ac{AUROC} (\ac{MAE} $= 0.102$, $n = 59$ variants; Fig.~\ref{fig:subset_prediction}c,d). Across all ten models, the median per-target \ac{MAE} was $0.060$ (\ac{TCR}) and $0.080$ (\ac{BCR}), higher than dataset-level \ac{MAE} (Fig.~\ref{fig:subset_prediction}g), reflecting the smaller sample sizes available per target.}

\edited{At per-target resolution, the distance-conditioned correction yielded the largest gains on \ac{TCR} epitope-split prediction (samples grouped by which epitope they target), the clinically relevant setting: median $|r| = 0.83$ with the full \ac{CaliPPer} versus $0.73$ with the density-ratio component alone (\ac{PAPE}~\cite{bialek2024pape}) and $0.69$ with the \ac{MCBPE}~\cite{bialek2024model} density-ratio variant (higher $|r|$ indicates closer agreement between predicted and actual performance). On distance-split subsets (samples grouped by S2DD distance from training), $|r| = 0.90$ vs $0.80$ vs $0.82$ (Fig.~\ref{fig:subset_prediction}h,l,p). On \ac{BCR} per-target prediction the distance correction added little ($|r| \approx 0.77$ for all three), reflecting a data-regime ceiling (compressed \ac{BCR} cross-test range; preceding section) rather than a method-specific limitation. The \ac{TCR} gain traces to the per-sample distance signal \ac{CaliPPer} adds on top of the density-ratio base, which density-ratio estimation alone cannot incorporate.}

\edited{On \ac{TCR}, per-target \ac{AUROC} prediction error showed a moderate negative association with the underlying model's \ac{AUROC} ($r = -0.53$; Fig.~\ref{fig:subset_prediction}e), indicating that prediction error grows on targets where the underlying model achieves lower \ac{AUROC}. Per-target heatmaps show that prediction error is generally lower on training-represented epitopes and variants than on rare or unseen ones (Fig.~\ref{fig:subset_prediction}i,k). We report \ac{AUROC} rather than \ac{AP} as the per-target indicator because per-target \ac{AP} is confounded by per-target class prevalence (see Discussion); this confound averages out at the dataset level.}

\begin{figure}[t]
    \centering
    \includegraphics[width=\linewidth]{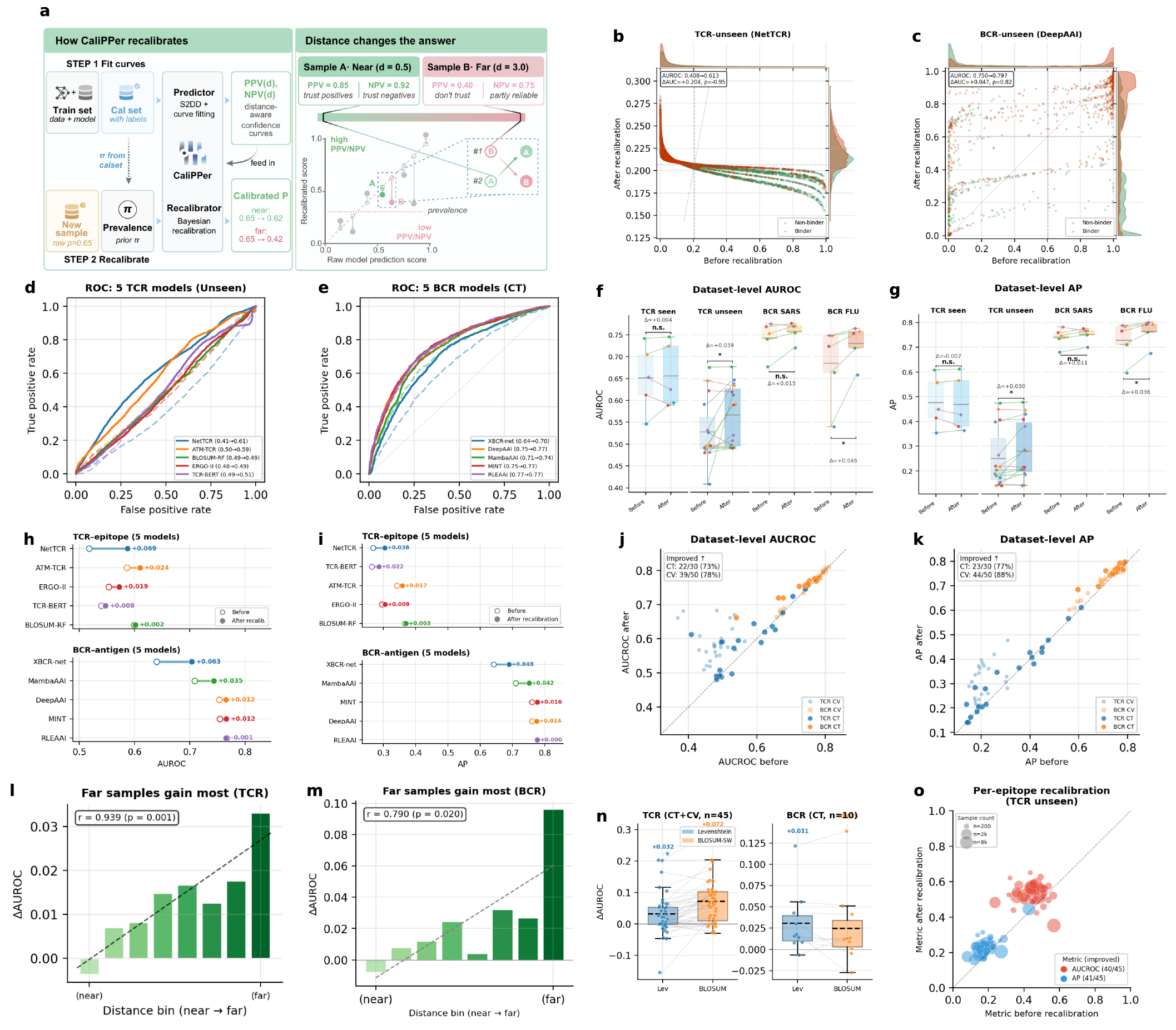}
    \caption{\edited{\textbf{CaliPPer recalibration improves prediction reliability across 10 models.} \textbf{a}, Conceptual overview of CaliPPer's two-step distance-aware Bayesian recalibration. \emph{Step 1}: CaliPPer fits distance-dependent PPV($d$), NPV($d$) confidence curves on a calibration set. \emph{Step 2}: a Bayesian recalibrator combines these curves with prevalence to map each prediction to a calibrated probability at its S2DD distance. PPV/NPV examples and recalibrated-vs-raw scatter illustrate the resulting rank flips that improve discrimination. \textbf{b, c}, Per-sample recalibrated vs.\ raw scores on unseen targets: TCR unseen epitopes (NetTCR, $\Delta$AUROC $= +0.204$; \textbf{b}) and BCR unseen SARS-CoV-2 variants (DeepAAI, $\Delta$AUROC $= +0.047$; \textbf{c}), coloured by true label; marginal KDEs show score redistribution. \textbf{d}, ROC curves for 5 TCR models on unseen epitopes (dashed: raw; solid: recalibrated); legend reports AUROC raw $\to$ recalibrated. \textbf{e}, Same as \textbf{d} for 5 BCR models on cross-test data. \textbf{f, g}, Paired boxplots of dataset-level AUROC and AP before and after recalibration, grouped by domain (TCR-seen, TCR-unseen, BCR-SARS-CoV-2, BCR-influenza); significance markers above each pair indicate one-sided Wilcoxon signed-rank test results (\textsuperscript{*}$p<0.05$, \textsuperscript{**}$p<0.01$; $n=5$ for TCR-seen and each BCR domain, $n=15$ for TCR-unseen; full statistics in Methods, ``Statistical analysis''). \textbf{h}, Per-model $\Delta$AUROC for all 10 models. Open circles, raw AUROC; filled circles, recalibrated AUROC; labels indicate $\Delta$. \textbf{i}, Same as \textbf{h} for $\Delta$AP. \textbf{j, k}, Dataset-level AUROC and AP before versus after recalibration across all model--test-set combinations: cross-test (solid markers) and cross-validation half-split (lighter markers). Points above the diagonal indicate improvement; annotations report the fraction of combinations improved (CT and CV). \textbf{l, m}, Per-distance-bin $\Delta$AUROC for TCR and BCR. Levenshtein-based S2DD; far-from-training samples gain most. Pearson correlation between distance bin and $\Delta$AUROC ($n=8$ bins): TCR $r=0.939$, $p=0.001$; BCR $r=0.790$, $p=0.020$ (95\% CIs in Methods). \textbf{n}, Levenshtein vs BLOSUM as base S2DD metric. Left: TCR (CT+CV, $n=45$); right: BCR (CT per-domain, $n=10$). BLOSUM outperforms Levenshtein for TCR ($+0.072$ vs $+0.032$ mean $\Delta$AUROC); equivalent gains for BCR. \textbf{o}, Per-epitope recalibrated vs.\ raw metric, TCR unseen epitopes: AUROC (red, 40/45 improved) and AP (blue, 41/45 improved); bubble size $\propto$ per-epitope sample count.}}
    \label{fig:recalibration}
\end{figure}

\subsection*{CaliPPer recalibration improves prediction reliability}

\edited{Beyond aggregated predictions, we sought to determine whether distance-aware Bayesian recalibration could improve per-sample prediction accuracy. The recalibration uses the learned \ac{S2DD} distance--performance relationship to adjust each prediction based on how far it lies from training. It operates entirely post-hoc without model retraining, additional experiments or architectural changes (Methods; Fig.~\ref{fig:recalibration}a). For representative \ac{TCR} and \ac{BCR} models, recalibration reshaped the score distribution to better separate binders from non-binders (Fig.~\ref{fig:recalibration}b,c). Recalibration per-sample extends label-free performance estimation beyond the aggregate-only metric estimates that density-ratio formulations like \ac{PAPE}~\cite{bialek2024pape} or \ac{MCBPE}~\cite{bialek2024model} return. It produces per-sample recalibrated confidence scores for downstream candidates ranking, an output type that the density-ratio paradigm cannot produce by design. The magnitude and direction of each adjustment were set jointly by distance-dependent \ac{PPV}($d$) and \ac{NPV}($d$), combined through the closed-form Bayesian update of Fig.~\ref{fig:overview}b. Where both predictive values exceeded their base rates, the local model was informative and recalibrated scores tracked the raw prediction; where they approached the base rates, scores were pulled toward the prior prevalence $\pi$; and where the local model became anti-informative (\ac{PPV}($d$) $+$ \ac{NPV}($d$) $< 1$), the update crossed $\pi$, mapping confident raw positives to recalibrated scores below $\pi$ and confident raw negatives above it (off-diagonal points crossing the prior line in Fig.~\ref{fig:recalibration}b).}

\edited{Across all ten models, recalibration improved discriminative performance, with ROC curves shifting upward for all five \ac{TCR} models on unseen epitopes and all five \ac{BCR} models on cross-test data (Fig.~\ref{fig:recalibration}d,e). Aggregate gains were consistent across two evaluation protocols, cross-test and cross-validation half-split. In the cross-validation half-split, each cross-validation fold is sorted by \ac{S2DD} distance and partitioned into a calibration half (used to fit \ac{PPV}/\ac{NPV} curves) and a held-out half (used for evaluation); this protocol simulates the within-study deployment scenario in which a user's labelled validation data and unlabelled query cohort are drawn from the same source and span a similar distance range. Under this protocol, 39/50 (78\%) and 44/50 (88\%) model--fold combinations improved \ac{AUROC} and \ac{AP} respectively, compared with 22/30 (73\%) and 23/30 (77\%) under cross-test (Fig.~\ref{fig:recalibration}f--k). Per-model mean \ac{AUROC} gains spanned $+0.002$ to $+0.069$ (\ac{TCR}) and $-0.001$ to $+0.063$ (\ac{BCR}; Fig.~\ref{fig:recalibration}h,i), while individual test-set gains were larger, reaching $+0.204$ for NetTCR on unseen epitopes (per-test-set breakdowns in Supplementary~Tables~S20,~S21). At near-distance bins, where raw model scores are most reliable, recalibration left predictions largely intact (TCR nearest-bin mean $\Delta$\ac{AUROC} ${\approx}-0.003$; Fig.~\ref{fig:recalibration}l), confining recalibration's downside to the regime where the underlying model is already reliable. This bounded downside is a built-in safety property of \ac{CaliPPer}'s distance-aware recalibration, which reduces to standard Platt scaling when the calibration data lack a non-flat \ac{PPV}($d$)/\ac{NPV}($d$) signal (Supplementary Note~7).}

\edited{This safety property (that CaliPPer's downside is bounded near that of standard recalibration) supports a practical pre-deployment check: users can run the same half-split protocol on their own labelled validation data, then measure $\Delta$\ac{AUROC} and $\Delta$\ac{AP} on the held-out half. A positive result confirms that the distance information is adding value beyond what a global rescaling could achieve. The two protocols correspond to two distinct deployment scenarios: the cross-validation half-split ($78\%/88\%$) reflects within-study deployment, where calibration and query data come from the same source and span similar distance ranges, the common case for binding prediction; cross-test ($73\%/77\%$) reflects across-study deployment, where the query is a fully external dataset, a stricter scenario. Improvements were largest on out-of-distribution data: \ac{TCR} unseen-epitope mean \ac{AUROC} gained $+0.066$ (up to $+0.204$ for NetTCR) and \ac{BCR} influenza \ac{AUROC} $+0.046$ (Fig.~\ref{fig:recalibration}f,g).}

\edited{The magnitude of recalibration gain scaled with distance from training. Per-distance-bin $\Delta$\ac{AUROC} correlated positively with \ac{S2DD} distance for both \ac{TCR} ($r = 0.939$, $p = 0.001$; Fig.~\ref{fig:recalibration}l) and \ac{BCR} ($r = 0.790$, $p = 0.020$; Fig.~\ref{fig:recalibration}m), with the largest gains on the most distant samples. The choice of base distance affected \ac{TCR} but not \ac{BCR} gains: BLOSUM yielded stronger \ac{TCR} improvements ($+0.072$ vs $+0.032$ mean $\Delta$\ac{AUROC} for Levenshtein), while both formulations gave equivalent gains for \ac{BCR} (Fig.~\ref{fig:recalibration}n; Methods, S2DD Summary describes the per-domain base-distance choice and sequence-length rationale). At the per-target level, 40 of 45 epitope subsets gained \ac{AUROC} and 41 of 45 gained \ac{AP} after recalibration (Fig.~\ref{fig:recalibration}o; per-test-set breakdowns in Supplementary Tables~S20,~S21). Recalibration thus delivers the largest gains where they matter most: on samples far from training, where raw model scores are least reliable.}

\begin{figure}[t]
    \centering
    \includegraphics[width=\linewidth]{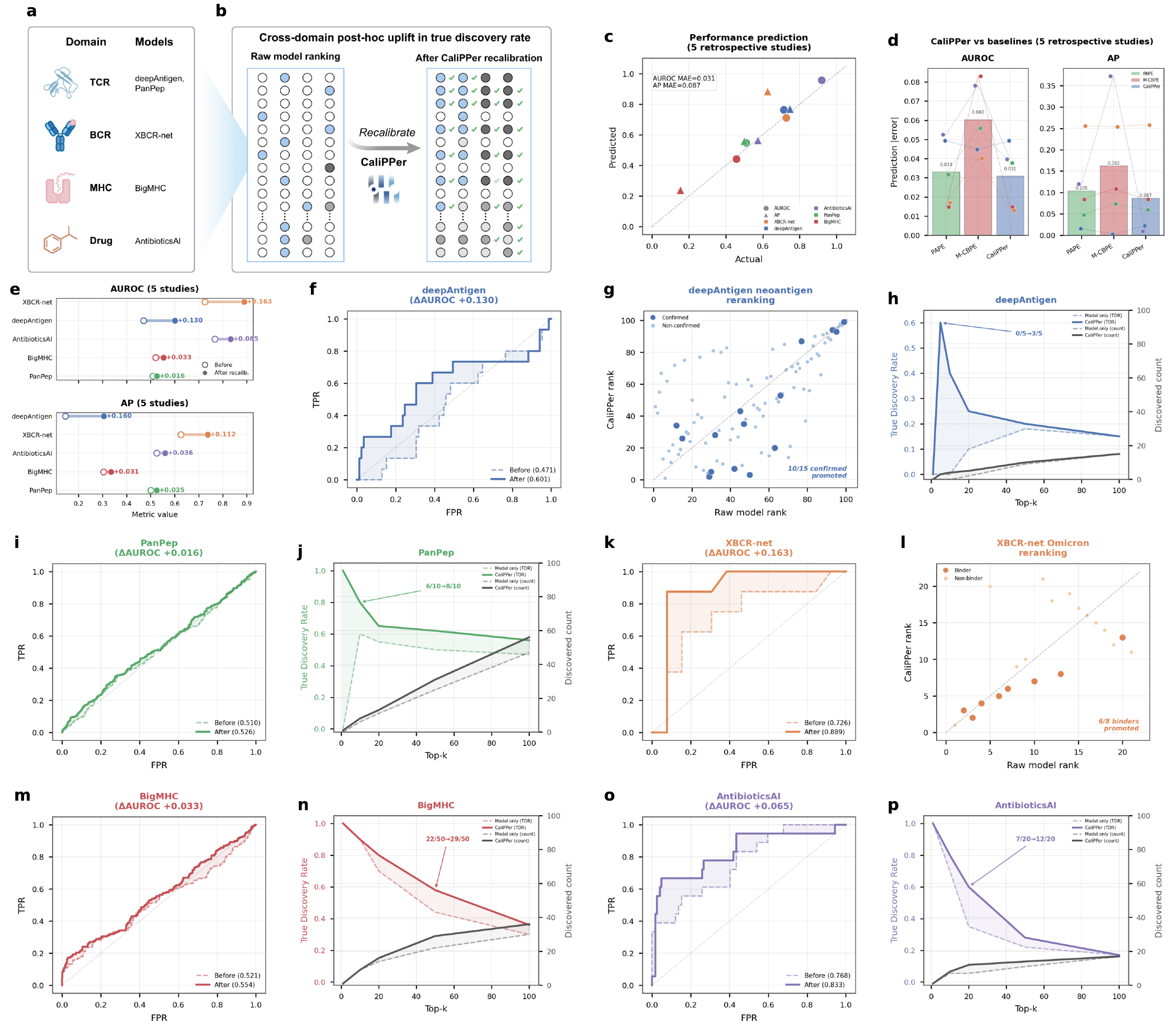}
    \caption{\edited{\textbf{Independent retrospective validation across 5 published studies spanning immunology and drug discovery.} \textbf{a}, Five retrospectively re-evaluated published models across four domains: deepAntigen, PanPep (\ac{TCR}--epitope), XBCR-net (\ac{BCR}--antigen), BigMHC (MHC--peptide) and AntibioticsAI (small-molecule). \textbf{b}, Schematic of cross-domain post-hoc true discovery rate (TDR) uplift: CaliPPer re-ranks raw model scores (left) to promote true positives (right), increasing top-$k$ TDR; per-study uplifts at study-specific cutoffs in inset. \textbf{c}, Performance prediction accuracy: predicted vs actual AUROC (circles) and AP (triangles) for each study (colour-coded). Points near the diagonal indicate accurate prediction. \textbf{d}, Performance prediction error decomposition: the density-ratio base estimate alone (PAPE; M-CBPE density-ratio variant shown for reference) versus the full CaliPPer (density-ratio base extended with S2DD-anchored distance-conditioned correction), with per-study errors (dots, connected by lines) for AUROC (left) and AP (right). \textbf{e}, Recalibration $\Delta$AUROC (top) and $\Delta$AP (bottom) across all 5 studies. Open circles $=$ before, filled $=$ after; all 5 studies show positive $\Delta$AUROC. \textbf{f--h}, deepAntigen: ROC curves before and after recalibration (\textbf{f}); neoantigen reranking scatter showing model rank vs CaliPPer rank for confirmed immunogenic (dark) and non-confirmed (light) candidates (\textbf{g}); and top-$k$ true discovery rate (TDR, left axis) with cumulative discovered count (right axis) (\textbf{h}). \textbf{i, j}, PanPep (zero-shot unseen peptides): ROC (\textbf{i}) and top-$k$ TDR with discovered count (\textbf{j}). \textbf{k, l}, XBCR-net: ROC (\textbf{k}) and Omicron-era antibody reranking scatter showing raw model rank vs CaliPPer rank for binders (dark) and non-binders (light) (\textbf{l}). \textbf{m, n}, BigMHC: ROC (\textbf{m}) and top-$k$ TDR with discovered count (\textbf{n}). \textbf{o, p}, AntibioticsAI: ROC (\textbf{o}) and top-$k$ TDR with discovered count (\textbf{p}), demonstrating generalisation beyond immunology to small-molecule drug discovery. See Supplementary Table~S1 for dataset and split details.}}
    \label{fig:retrospective}
\end{figure}

\subsection*{Independent retrospective validation across five published studies}

\edited{To test whether \ac{CaliPPer} could be applied to published models as a post-hoc layer, we deployed it retrospectively (using each study's released predictions, with no retraining, fine-tuning or architecture modification) to five models spanning immunology and drug discovery: deepAntigen~\cite{que2025deepantigen} (\ac{TCR} neoantigen binding), PanPep~\cite{gao2023panpep} (pan-allele \ac{TCR} meta-learning), XBCR-net~\cite{xbcrnet} (\ac{BCR}--antigen binding), BigMHC~\cite{albert2024bigmhc} (MHC-I immunogenicity) and AntibioticsAI~\cite{wong2024antibiotics} (small-molecule antibiotic activity). Calibration data and evaluation data splits for each study are detailed in Methods (Supplementary Table~S1).}

\edited{CaliPPer's dataset-level predictions, which combine the density-ratio base estimate with \ac{S2DD}-anchored distance correction, consistently achieved lower mean absolute error than \ac{PAPE} ($0.087$ vs $0.105$ for \ac{AP}; $0.031$ vs $0.033$ for \ac{AUROC}) and \ac{MCBPE} ($0.162$ \ac{AP}; $0.060$ \ac{AUROC}), averaged across the five retrospective studies (Fig.~\ref{fig:retrospective}c,d). The \ac{AUROC} margin is modest while the \ac{AP} margin is larger ($0.018$ over PAPE, $0.075$ over M-CBPE) and is the more informative metric here, because \ac{AP} responds more strongly to the prevalence shifts that distance-aware correction targets. \ac{CaliPPer} leads on the averaged metrics across all five retrospective studies, including the small-molecule case where sequence-based density ratios do not apply. Bayesian recalibration improved both \ac{AUROC} and \ac{AP} in all five studies (Fig.~\ref{fig:retrospective}e): \ac{AUROC} gains ranged from $+0.016$ (PanPep, the zero-shot setting) to $+0.163$ (XBCR-net), and \ac{AP} gains from $+0.025$ to $+0.160$ (per-study breakdown in Supplementary Table~S22).}

\edited{These recalibration gains directly translate into therapeutic discoveries. For deepAntigen~\cite{que2025deepantigen}, performance prediction (Fig.~\ref{fig:retrospective}c,d) was evaluated on the independent ImmuneCODE dataset~\cite{nolan2023pird} ({$\sim$}50{,}000 \ac{TCR}--epitope pairs), whereas recalibration (Fig.~\ref{fig:retrospective}f,g,h) was evaluated on 100 clinical neoantigens from 5 cancer patients with ELISPOT-confirmed immunogenicity labels. This distinction reflects study design~\cite{que2025deepantigen}: the 100 neoantigens were pre-selected by deepAntigen as top-ranked candidates from pools of 1{,}167--41{,}606 per patient, making them unsuitable as an independent test set for performance prediction but representing the realistic scenario for recalibration: a researcher screens candidates computationally, then applies \ac{CaliPPer} to re-rank the shortlist before experimental validation. Under this protocol, none of the model's top-5 neoantigen candidates were confirmed immunogenic (\ac{TDR} $= 0/5$); after recalibration, 3 of the top 5 were confirmed (\ac{TDR} $= 3/5$), and 10 of 15 confirmed immunogenic peptides were promoted in rankings (Fig.~\ref{fig:retrospective}g,h). The same five-candidate experimental budget thus yielded three actionable neoantigen targets instead of zero. For XBCR-net, recalibration improved TDR at $k{=}8$ from 5/8 to 7/8, correctly promoting 6 of 8 known Omicron-era binders while demoting non-binders that the raw model had ranked highly (Fig.~\ref{fig:retrospective}k,l). PanPep, evaluated on 491 zero-shot peptides with 0\% training overlap, showed positive gains in both metrics ($\Delta$\ac{AUROC} $= +0.016$, $\Delta$\ac{AP} $= +0.025$) and improved Top-10 TDR from 6/10 to 8/10 (Fig.~\ref{fig:retrospective}i,j), confirming that recalibration adds value in this zero-shot regime. For BigMHC, true cross-dataset evaluation (688-sample held-out validation set as calibration, 834-sample independent MANAFEST cohort as test) improved both discrimination ($\Delta$\ac{AUROC} $= +0.033$, $\Delta$\ac{AP} $= +0.031$) and true discovery rates (TDR at $k{=}50$: 22/50 $\to$ 29/50; Fig.~\ref{fig:retrospective}m,n). AntibioticsAI, which uses Morgan fingerprint distances rather than sequence-based metrics, achieved $+0.065$ \ac{AUROC} and $+0.036$ \ac{AP} improvements with TDR at $k{=}20$ improving from 7/20 to 12/20 (Fig.~\ref{fig:retrospective}o,p).}

\edited{Across all five studies, the same mechanism produced the gains: CaliPPer's distance-dependent recalibration re-ranks candidates within each distance regime according to the local \ac{PPV}/\ac{NPV} signal, promoting reliable predictions and demoting overconfident ones. The largest gains occurred on the most distant samples, both in the primary benchmarks (Fig.~\ref{fig:recalibration}l,m; $r = 0.939$ TCR, $r = 0.790$ BCR) and in the retrospective top-$k$ outcomes (deepAntigen $0/5 \to 3/5$ confirmed neoantigens, Fig.~\ref{fig:retrospective}g,h; XBCR-net Omicron binders $5/8 \to 7/8$, Fig.~\ref{fig:retrospective}k,l). The modular design of the S2DD distance, which supports Levenshtein, BLOSUM substitution matrices and Morgan fingerprints, enables this cross-domain generality without modifying the prediction or recalibration pipelines (robustness analyses across distance formulations, metric selection and hyperparameter sensitivity in Supplementary Notes). \ac{CaliPPer} therefore generalises across receptor families, prediction tasks and substrate types without modification.}


\section*{Discussion}

\edited{CaliPPer is, to our knowledge, the first framework that converts the distance--performance relationship in immune-receptor binding into a quantitative performance estimate and per-sample recalibration, operating at three progressively finer resolutions. At the model level, the \ac{S2DD} degradation curve provides a distance-resolved generalisability profile: model developers can compare architectures not only by aggregate accuracy but by examining at which distance ranges each model excels, selecting the most suitable model for a given deployment scenario based on the expected distributional shift. At the data level, CaliPPer's distance--performance curves predict aggregate performance on new cohorts without labels; when cohorts can be split by target (epitope, antigen variant or protein family), per-subset predictions provide graded confidence estimates, allowing researchers to identify which targets a model can handle reliably and which require additional experimental validation. At the sample level, CaliPPer's Bayesian recalibration assigns per-prediction confidence: screening teams can triage candidates by testing high-confidence predictions first and flagging low-confidence predictions as potential false negatives. Together, these three levels insert a quantitative decision layer between computational prediction and experimental validation.}

\edited{Three components distinguish \ac{CaliPPer} from prior label-free estimators. First, the multi-chain \ac{S2DD} distance combines per-chain similarities into a single distance from each new sample to the training distribution; the chain weights are learned automatically from the training data (so the user does not pick which chain matters most), and the underlying similarity can be sequence edit distance, substitution-matrix score, learned embedding, or molecular-fingerprint similarity, supporting the same pipeline across receptor sequences and small-molecule fingerprints. Second, a performance predictor that combines a label-free density-ratio per-sample base estimate (in the spirit of \ac{PAPE}~\cite{bialek2024pape}) with a distance-anchored correction fitted on \ac{S2DD}, recovering accuracy on the high-shift samples where density-ratio estimation alone fails (Supplementary Note). Third, a Bayesian recalibration that uses the calibration set's positive and negative predictive values (\ac{PPV}, \ac{NPV}) at each distance to assign every new prediction a confidence score scaled to how far it lies from training. Unlike standard post-hoc rescaling methods (Platt scaling, isotonic regression), which preserve the original ranking and therefore can only improve calibration error, this distance-dependent adjustment can also rerank candidates. To our knowledge, this is the only post-hoc recalibration of immune-receptor binding predictions for which this property holds without retraining the model. The method automatically adjusts its behaviour based on the model's local reliability: when scores are already accurate, it behaves like a simple rescaling; when scores are unreliable, it leans on the distance information (Supplementary Note~9).}

\edited{These results carry practical implications for both model development and training-data design. For model developers, we recommend reporting degradation curves alongside aggregate performance metrics in benchmarking studies, providing a richer characterisation than point estimates alone. For benchmarking efforts such as Lu et al.~\cite{lu2025assessment} (50 \ac{TCR}--epitope models), \ac{CaliPPer} transforms the observation that sequence diversity alone cannot predict model failure into a working framework, enabling developers to rank models by how performance varies with distributional distance and to select those with flatter curves for diverse-target deployment. For training-data design, the $\sigma \!\cdot\! C$ chain analysis provides a per-dataset, hyperparameter-free readout of which sequence dimension is most concentrated in a given training distribution; recomputing it on a new training set indicates where added diversity would most reduce out-of-distribution distance, supporting data-collection planning.}

\edited{For clinical translation, CaliPPer's recalibration directly addresses the cost-effectiveness of computational screening. In the deepAntigen neoantigen pipeline, 0 of the top 5 raw-score candidates were confirmed immunogenic; after \ac{CaliPPer} recalibration, 3 of the top 5 were confirmed, with \ac{TDR} gains across all five retrospective studies at clinically relevant cutoffs (per-study cutoff ranges in Supplementary~Table~S22). At the top of the ranking, this translates to roughly 1--2 additional confirmed hits per 10 candidates tested, by replacing overconfident false positives with genuine binders that the raw model had ranked lower. In antibody discovery and TCR-based immunotherapy pipelines, where binding validation requires days to weeks per candidate~\citemethods{lu2021deep,hudson2023predict}, this translates to substantial savings in time and reagents.}

\edited{The framework is regime-invariant across diverse evaluation contexts: validation spans ten models from eight architectural families, covering \ac{TCR}--epitope binding (783 epitopes from diverse pathogens) and \ac{BCR}--antigen binding (SARS-CoV-2 and influenza; 80 antigen variants), and is further extended through retrospective application to five published studies spanning \ac{TCR} neoantigen, \ac{BCR} antigen, MHC--peptide and small-molecule binding tasks (Fig.~\ref{fig:retrospective}). The modular distance design supports Levenshtein, BLOSUM, ESM-2 and RMSD variants, suggesting applicability to any molecular interaction with multiple sequence components, including MHC--peptide binding, protein--protein interactions and drug--target affinity.}

\edited{CaliPPer's degradation, prediction and recalibration patterns hold across a 20-percentage-point gap in absolute model performance between \ac{TCR} (median \ac{AUROC} ${\approx}$0.55) and \ac{BCR} (${\approx}$0.75) regimes, which differ both in data conventions (unseen-epitope vs unseen-variant generalisation) and in metric dynamic range (Supplementary Notes~4, 5). This regime-invariance suggests the framework characterises a distance--performance relationship intrinsic to binding-prediction problems rather than a property of any particular dataset or evaluation protocol. The TCR ${\approx}$0.55 figure includes the hardest unseen-epitope splits where models approach chance-level \ac{AUROC}; \ac{CaliPPer}'s predicted values track these low actual values (Fig.~\ref{fig:prediction}e) rather than defaulting to $0.5$.}

\edited{Three considerations qualify CaliPPer's applicability. First, \ac{CaliPPer} requires at least $30$ calibration samples per distance bin (with at least $8$ samples per bin from the smaller class, whether binders or non-binders) to fit reliable curves; very small calibration cohorts or those with substantial class imbalance may reduce accuracy. Comparable sample-size floors apply to density-ratio estimators (\ac{PAPE}~\cite{bialek2024pape}: ${\sim}6{,}000$ reference samples and ${\sim}2{,}000$ per chunk; \ac{MCBPE}~\cite{bialek2024model}: ${\sim}300$--$600$ per chunk), which similarly degrade in sparse data regimes. Second, \ac{CaliPPer} presumes the calibration cohort spans a distance range that brackets the query; when query distances exceed the calibration maximum, the fitted distance--performance curve extrapolates and prediction reliability decreases. Third, recalibration gains depend on the calibration cohort exhibiting a non-flat \ac{PPV}($d$)/\ac{NPV}($d$) profile representative of the deployment regime; when this condition is weakly met, as in the PanPep zero-shot setting, where the calibration cohort and test cohort share no peptides ($\Delta$\ac{AUROC} $= +0.016$, the smallest of the five retrospective studies), \ac{CaliPPer} reduces to standard Platt scaling and gains are correspondingly modest, with the cal-half $\Delta$\ac{AUROC} diagnostic (Fig.~\ref{fig:recalibration}f--k; further detail at the start of the next paragraph) providing a pre-deployment check.}

\edited{Two deployment considerations merit emphasis. CaliPPer's recalibration reuses any labelled validation cohort already available from model development, provided it contains targets not present in training (the standard requirement for held-out validation). It adds a distance dimension to the rescaling that standard post-hoc probability-adjustment methods (such as Platt scaling or isotonic regression, which apply the same adjustment function to all predictions) would perform globally. For all five retrospective studies, the calibration set was drawn from cohorts the original authors had already published (Supplementary Table~S1), confirming that pre-existing validation data is sufficient when it covers novel targets. The minimum-sample requirement ($\geq 30$ per bin) applies to the calibration cohort, not to the query: per-sample recalibration and per-model comparison via the fitted distance--performance curve then apply to any number of query candidates, including small cohorts (e.g., 5--10 TCRs or antigens), provided their distances fall within the range spanned by the calibration cohort.}

\edited{Three complementary diagnostics support deployment. For performance prediction, comparing the calibration-data distance range to the target-data range (Results; Supplementary Note~6) indicates whether to trust the correlation $R$ or rely on \ac{MAE} alone. For recalibration, computing the change in \ac{AUROC} on a half-split of the labelled validation data (the protocol described in our cross-validation evaluation, Fig.~\ref{fig:recalibration}f--k) provides a pre-deployment signal: a positive improvement indicates that the \ac{PPV} and \ac{NPV} vary with distance in the calibration cohort and that \ac{CaliPPer}'s distance-aware adjustment will alter predictions relative to a global rescaling. As a basic prerequisite, the calibration cohort must support fitting a non-flat distance--performance curve. This fails either (i) when the cohort is too concentrated in one distance regime to fit any curve, or (ii) when it lies entirely in the flat plateau region of the curve (cf.\ Fig.~\ref{fig:degradation}c,d for the shape), as in the IEDB non-SARS pathogens (Supplementary Table~S16); in case (ii), the model's cal-set \ac{AUROC} already reveals it is at chance, so CaliPPer's prediction is not needed. In both cases the distance-aware recalibration reduces to standard Platt scaling and the framework correctly returns the model's predictions unchanged.}

\edited{Future extensions include integration with model-internal uncertainty estimates (Monte Carlo dropout, ensembles) for complementary confidence signals, CaliPPer-guided active learning for data acquisition, training data sufficiency estimation via slope saturation analysis, extension to continuous affinity prediction ($K_\mathrm{d}$, IC$_{50}$), and combining distance types for richer distributional characterisation, particularly evaluating whether ESM-2 or RMSD distances recover similarity between structurally related but sequence-distant chains where Levenshtein alone underestimates it. Although our TCR data is predominantly HLA class~I-restricted (reflecting the natural ${\sim}$90/10 class~I:II composition of public TCR databases such as VDJdb and McPAS-TCR), a dedicated benchmark for HLA class~II--restricted CD4$^{+}$ \ac{TCR}--peptide binding remains a natural next direction; class~II peptides span a wider length range (12--25 amino acids) over which the data-driven $\sigma \!\cdot\! C$ chain weighting would auto-adapt, and the S2DD framework itself does not depend on MHC class and is directly applicable given a class~II--trained underlying model. The framework and all model predictions are available as an open-source package to enable immediate adoption. By making distributional distance a first-class evaluation dimension, \ac{CaliPPer} reframes binding-prediction reliability from a property of the model to a property of the model--test-distribution pair, with implications for benchmarking practice, deployment confidence and experimental prioritisation.}

\begin{spacing}{0.05}
    \bibliography{ref}   %
  \end{spacing}
\clearpage

\section*{Methods}
\label{sec:method}

\subsection*{Framework Design}
\label{sec:framework_design}

\edited{CaliPPer (Calibration and Prediction of Performance) is a framework for evaluating, predicting and improving the reliability of binary binding-prediction models. The framework integrates two methodological components: (i)~a multi-chain distance metric, Sample-to-Domain Distance (\ac{S2DD}), which measures the distributional distance between each test sample and the training distribution; and (ii)~a distance-aware Bayesian recalibration that adjusts per-sample predictions based on this distance. \ac{S2DD} is built on a base per-pair distance, LogDist (a log-transformed sequence similarity), aggregated across receptor and antigen chains via concentration-based weighting, then reduced to a scalar per sample via top-$K$ averaging. The Bayesian recalibration derives Platt-scaling parameters analytically from distance-dependent positive and negative predictive values (PPV and NPV) estimated on a labelled calibration set, eliminating the need for retraining the underlying model or fitting the Platt parameters via maximum likelihood.}

\edited{Together, these two components produce outputs at three resolution levels: a model-level generalisability score (the \ac{S2DD}--performance degradation slope), data-level prediction of aggregate performance on unlabelled test datasets (via bias-corrected decay curves), and sample-level calibrated binding probabilities (via Bayesian recalibration). All three outputs share the same underlying \ac{S2DD} distance computation and operate entirely post-hoc on any trained binary classifier.}

\subsection*{Problem Formulation}
\label{sec:problem_formulation}

\edited{To formalise the framework, we first establish notation and the three estimation tasks it addresses. Let $\mathcal{R} = \{r_1, r_2, \ldots, r_N\}$ denote the training (reference) set and $\mathcal{Q} = \{q_1, q_2, \ldots, q_M\}$ denote a new test (query) set. Each sample consists of $C$ chains or molecular components: for TCR data, $C = 3$ (peptide, CDR3$\alpha$, CDR3$\beta$); for BCR data, $C = 3$ (heavy chain, light chain, antigen variant); for small-molecule applications, the chain decomposition is replaced by molecular fingerprint feature vectors. A prediction model $\mathcal{M}$ trained on $\mathcal{R}$ produces a binding probability $\hat{y}_j = \mathcal{M}(q_j)$ for each query sample $q_j \in \mathcal{Q}$. Each query sample has an unobserved ground-truth label $y_j \in \{0, 1\}$.}

\edited{We address three estimation tasks simultaneously.}

\paragraph{Task 1: Distance function.} \edited{We seek a distance function $D(q \mid \mathcal{R})$ that maps each query sample to a scalar measuring its distance from the training distribution. We require two properties: (i)~\textbf{negative correlation}, i.e., for groups of samples binned by distance, the Pearson correlation between mean bin distance and bin-level performance is negative; and (ii)~\textbf{consistency}, i.e., the distance--performance relationship generalises across test datasets, data splits and prediction models, enabling reliable estimation on unlabelled data.}

\paragraph{Task 2: Performance prediction.} \edited{Given $\mathcal{R}$, $\mathcal{Q}$, $\mathcal{M}$ and the distances $\{D(q_j \mid \mathcal{R})\}_{j=1}^{M}$, predict the aggregate performance $\hat{\pi}(\mathcal{Q})$ of $\mathcal{M}$ on $\mathcal{Q}$ (measured by AUROC, AP or F1) without access to the ground-truth labels $\{y_j\}$. Accuracy is measured by mean absolute error against the true performance $\pi(\mathcal{Q})$ across multiple held-out test settings.}

\paragraph{Task 3: Sample-level recalibration.} \edited{For each query sample $q_j$ with raw prediction $\hat{y}_j$ and distance $D(q_j \mid \mathcal{R})$, produce a recalibrated probability $\tilde{y}_j$ that reflects both the raw model output and the sample's \ac{S2DD} distance from the training distribution, such that $\tilde{y}_j$ better approximates the true posterior $P(y_j = 1 \mid q_j)$. Quality is measured by improvement in discrimination (AUROC, AP) against the raw predictions $\hat{y}_j$.}

\edited{The subsequent sections describe the \ac{S2DD} distance metric, the binned evaluation and degradation analysis, the performance prediction algorithm and the Bayesian recalibration procedure.}

\begin{algorithm}[!h]
\caption{S2DD: Sample-to-Domain Distance}
\label{alg:s2dd}
\KwIn{Query sample $q$ with $C$ chains; training set $\mathcal{R} = \{r_1, \ldots, r_N\}$; parameters $k$, $b$, $K$}
\KwOut{Sample-to-domain distance $D(q)$}
\smallskip\hrule\smallskip
\tcp{Preprocessing (computed once from $\mathcal{R}$)}
\For{each chain $i = 1, \ldots, C$}{
  Compute $\sigma_i$ (LogDist std) and $C_i$ (Simpson's index, Eq.~\eqref{eq:simpson}) from $\mathcal{R}$\;
  $w_i \leftarrow \sigma_i \cdot C_i \,/\, \sum_{j} \sigma_j \cdot C_j$ \tcp*{Chain weight (Eq.~\eqref{eq:sigma_c})}
  $\tilde{w}_i \leftarrow w_i \cdot C$ \tcp*{Rescale to unit mean}
  Compute $\mu_i$, $\sigma_i^{d}$ from pairwise distances within $\mathcal{R}$ for chain $i$\;
}

\tcp{Level 0: Chain-to-chain distances}
\For{each reference $r_j \in \mathcal{R}$}{
  \For{each chain $i = 1, \ldots, C$}{
    $d_i \leftarrow \log\bigl(k \cdot (1 - L(q_i, r_{j,i}) + b)\bigr)$ \tcp*{Eq.~\eqref{eq:per_pair}}
    $z_i \leftarrow (d_i - \mu_i) \,/\, \sigma_i^{d}$ \tcp*{$z$-normalize}
  }
  \tcp{Level 1: Sample-to-sample distance}
  $j^{*} \leftarrow \operatorname{arg\,max}_{i}\; \tilde{w}_i \, z_i$ \tcp*{Weighted chain selection}
  $D(q, r_j) \leftarrow z_{j^{*}}$ \tcp*{Eq.~\eqref{eq:weighted_max_znorm}}
}

\tcp{Level 2: Sample-to-domain distance}
$\mathcal{N}_K \leftarrow$ indices of $K$ smallest values in $\{D(q, r_j)\}_{j=1}^{N}$\;
$D(q) \leftarrow \frac{1}{K} \sum_{j \in \mathcal{N}_K} D(q, r_j)$ \tcp*{Top-$K$ reduction (Eq.~\eqref{eq:topk})}
\Return{$D(q)$}
\end{algorithm}

\subsection*{\ac{S2DD}}
\label{sec:s2dd}

\edited{We introduce \ac{S2DD}, a metric designed for multi-chain receptor data that integrates per-chain distance computation, automatic chain weighting and sample-to-domain aggregation in a single framework.} \ac{S2DD} operates in three stages: (i)~per-pair distance computation for each chain, (ii)~multi-chain combination that merges per-chain distances into a single distance per query-reference sample pair, and (iii)~top-$K$ reduction across references to yield a scalar distance per query sample.

\subsubsection*{Per-Pair Distance}
\label{sec:per_pair}

For a query sequence $q$ and a reference sequence $r$, the per-pair distance is
\begin{equation}
\label{eq:per_pair}
    d(q, r) = \begin{cases}
        \log\bigl(k \cdot (1 - \mathrm{sim}(q, r) + b)\bigr) & \text{(Levenshtein, ESM-2; $k = 0.1,\, b = 0.1$)} \\
        \sqrt{\max\bigl(1 - \mathrm{sim}(q, r),\, 0\bigr)} & \text{(BLOSUM-SW, Morgan-fingerprint Tanimoto)}
    \end{cases}
\end{equation}
where $\mathrm{sim}(q, r) \in [0, 1]$ is a normalised base similarity. The transformation stabilises variance across the distance range: logarithmic compression expands the sparse near-training (high-similarity) tail of right-skewed raw distances, sharpening discrimination where degradation is most informative, while the square-root is appropriate for raw distances that are more evenly populated across $[0, 1]$ and avoids the $\log(0)$ singularity at $\mathrm{sim} = 1$. Per-base similarity definitions and ablations are in Supplementary Note~6.

\subsubsection*{Multi-Chain Combination}
\label{sec:multichain}

Immune receptors consist of multiple amino-acid chains. For a TCR sample, $C = 3$ chains are available: the peptide (epitope), CDR3$\alpha$, and CDR3$\beta$. For BCR data, $C = 3$ chains are similarly used: the heavy chain, light chain, and antigen variant sequence. Given pairwise log-distances $d_i(q, r)$ computed via Eq.~\eqref{eq:per_pair} for each chain $i \in \{1, \ldots, C\}$ between a query sample $q$ and a reference sample $r$, we combine the $C$ per-chain distances into a single multi-chain distance $D(q, r)$.

We first $z$-normalize each per-chain distance using training-set statistics: $z_i = (d_i(q, r) - \mu_i) / \sigma_i^{d}$, where $\mu_i$ and $\sigma_i^{d}$ are the mean and standard deviation of per-pair raw log-distances (Eq.~\eqref{eq:per_pair}) for chain~$i$, computed from a random subsample of training sequences ($\leq$500) evaluated against all training sequences, without top-$K$ reduction. These z-norm statistics $(\mu_i, \sigma_i^{d})$ are distinct from the chain-weight statistic $\sigma_i$ in Eq.~\eqref{eq:sigma_c}, which is the standard deviation of top-$K$-reduced per-sample LogDist scores. The combination function uses chain weights $\mathbf{w} = (w_1, \ldots, w_C)$ as \emph{selectors} rather than magnitude scalers. The weights are rescaled to have unit mean, $\tilde{w}_i = w_i \cdot C$, and the combined distance is the $z$-score of the chain selected by the largest weighted $z$-score:
\begin{equation}
\label{eq:weighted_max_znorm}
    j^{*} = \operatorname*{arg\,max}_{i}\; \tilde{w}_i \, z_i, \qquad D(q, r) = z_{j^{*}}.
\end{equation}
The output is a pure, unscaled $z$-score: weights influence only \emph{which} chain is selected, not the magnitude of the resulting distance. The combined distance is large whenever any sufficiently weighted chain is far from the training distribution, ensuring consistent degradation even when individual chains exhibit mixed signals.

\subsubsection*{Chain Weighting}
\label{sec:chain_weights}

The chain weights $\mathbf{w}$ are determined automatically from the training data using a hyperparameter-free formula that combines distance variability with sequence concentration. For each chain $i$, we compute (i)~$\sigma_i$, the standard deviation of LogDist values computed over a subsample of training sequences evaluated against the full training set, capturing the variability of the distance landscape for chain~$i$; and (ii)~$C_i$, Simpson's concentration index,
\begin{equation}
\label{eq:simpson}
    C_i = \sum_{s \in \mathcal{U}_i} P_i(s)^2,
\end{equation}
where $\mathcal{U}_i$ is the set of unique sequences for chain~$i$ and $P_i(s) = \mathrm{freq}(s) / N$. $C_i$ measures the probability that two randomly drawn sequences from the training data are identical: chains with concentrated distributions have large $C_i$, while chains with diverse, uniform distributions have $C_i \to 0$.

The chain weights are then:
\begin{equation}
\label{eq:sigma_c}
    w_i = \frac{\sigma_i \cdot C_i}{\displaystyle\sum_{j=1}^{C} \sigma_j \cdot C_j}.
\end{equation}
This formula upweights chains with concentrated training distributions, which are expected to be the most informative for distinguishing in-distribution from out-of-distribution queries. \edited{The chain weights are computed from each training set, not set by the user, so their values depend on the training distribution and on the trained model the framework is applied to. On the TCR datasets and models considered here, the peptide chain emerges as dominant ($C_{\mathrm{peptide}} \gg C_{\mathrm{CDR3}\alpha}, C_{\mathrm{CDR3}\beta}$); on the BCR datasets, the antigen-variant chain emerges as dominant. These dominance patterns reflect the training distributions used here, not biological properties of TCR--epitope or BCR--antigen binding, and we do not generalise them to untested settings; per-dataset numerical weights appear in the Supplementary Information.} A query with a novel epitope or antigen variant will produce a large distance on the concentrated chain, and the high weight ensures this signal drives the chain selection in Eq.~\eqref{eq:weighted_max_znorm}.

\subsubsection*{Sample-to-Distribution Reduction}
\label{sec:sample_to_dist}

After combining per-chain distances into a multi-chain distance $D(q, r_j)$ for each reference $r_j$, we reduce the $N$ combined distances to a scalar $D(q)$ by averaging over the $K$ nearest references:
\begin{equation}
\label{eq:topk}
    D(q) = \frac{1}{K} \sum_{j \in \mathcal{N}_K^{D}(q)} D(q, r_j),
\end{equation}
where $\mathcal{N}_K^{D}(q)$ denotes the indices of the $K$ references with the smallest combined distances. We use $K = 50$ for TCR data and $K = 30$ for BCR data. The top-$K$ selection identifies training samples that are jointly close to $q$ across all chains.

\subsubsection*{\ac{S2DD} Summary}
\label{sec:multichain_summary}

Combining the above, the \ac{S2DD} multi-chain distance is:
\begin{equation}
\label{eq:multichain}
    D(q) = \frac{1}{K} \sum_{j \in \mathcal{N}_K^{D}(q)} z_{j^{*}(q, r_j)},
\end{equation}
where $j^{*}(q, r_j)$ selects the chain with the largest $\tilde{w}_i z_i$ for each reference $r_j$ (Eq.~\eqref{eq:weighted_max_znorm}), and $\mathcal{N}_K^{D}(q)$ selects the $K$ references with the smallest combined distances.

Equation~\eqref{eq:multichain} makes explicit the three-level aggregation hierarchy that constitutes the core technical design of S2DD. At the base level (\emph{chain-to-chain}), the log-transformed Levenshtein ratio (Eq.~\eqref{eq:per_pair}) computes a distance between a single query chain and a single reference chain. At the first aggregation level (\emph{sample-to-sample}), the $\sigma \!\cdot\! C$ weighting (Eq.~\eqref{eq:sigma_c}) and weighted-max $z$-norm combination (Eq.~\eqref{eq:weighted_max_znorm}) merge the $C$ per-chain distances into a single distance $D(q, r)$ between a query-reference pair. At the second aggregation level (\emph{sample-to-domain}), the top-$K$ reduction (Eq.~\eqref{eq:topk}) collapses $N$ sample-to-sample distances into a scalar $D(q)$ measuring the query's distance from the entire training distribution. Prior distance-based analyses of immune receptor generalisation~\cite{castorina2025assessing} operate only at the chain-to-chain level; our framework provides the complete pipeline from individual chain comparisons to a per-sample distributional distance. Importantly, the base distance metric is a modular component: the log-transformed Levenshtein ratio (LogDist) can be replaced with BLOSUM substitution-matrix distances~\cite{henikoff1992amino}, C$\alpha$-RMSD derived from predicted structures via ESMFold~\cite{lin2023esm2} or AlphaFold~\cite{jumper2021highly}, or learned embedding distances, without modifying the aggregation levels above it. The choice of base distance depends on the sequence-length regime: BLOSUM (square-root-transformed Smith--Waterman BLOSUM62 similarity) is preferred for short sequences ($\leq$30 amino acids; TCR epitopes, CDR3 loops, MHC pseudo-sequences), where substitution-matrix similarity captures biochemically meaningful variation; Levenshtein is preferred for long sequences ($\geq$100 amino acids; BCR heavy and light chains and viral RBD variants), where BLOSUM alignment scores saturate and compress the distance range. For sequences in the intermediate range (30--100 amino acids), which were not present in our primary evaluation, we recommend comparing both formulations on a held-out calibration set. Unless otherwise noted, all TCR results in this study use BLOSUM S2DD and all BCR results use Levenshtein S2DD.

Algorithm~\ref{alg:s2dd} summarizes the complete \ac{S2DD} computation.

\subsection*{Binned Evaluation and Correlation Analysis}
\label{sec:binned_eval}

\edited{To analyse how model performance varies with distance, we partition the query set into $B$ equal-sized bins ordered by increasing distance, using an adaptive bin count $B = \max(4,\, \min(8,\, \lfloor n_{\mathrm{min}} / 8 \rfloor))$ to balance resolution against statistical power, where $n_{\mathrm{min}}$ is the minority-class count.} Specifically, we sort $\mathcal{Q}$ by $D(q|\mathcal{R})$ and divide the sorted samples into $B$ consecutive groups of approximately $\lfloor M/B \rfloor$ samples each (the last bin absorbs any remainder).

For each bin $b \in \{1, \ldots, B\}$, we compute the mean distance $\bar{d}_b$ and a performance metric $\pi_b$ (e.g., AUROC, AP or F1) evaluated on the samples within that bin. We then assess the degradation trend by computing the Pearson correlation coefficient between the bin distances and bin performances:
\begin{equation}
\label{eq:pearson}
    r = \frac{\sum_{b=1}^{B} (\bar{d}_b - \bar{d})(\pi_b - \bar{\pi})}{\sqrt{\sum_{b=1}^{B} (\bar{d}_b - \bar{d})^2 \sum_{b=1}^{B} (\pi_b - \bar{\pi})^2}},
\end{equation}
where $\bar{d}$ and $\bar{\pi}$ are the means across bins. A strongly negative $r$ (with $p < 0.05$) indicates that model performance tends to degrade with increasing distance from the training distribution, confirming the expected negative correlation.

To assess the robustness of this relationship beyond linear assumptions, we additionally compute the Spearman rank correlation coefficient $\rho$ between the bin distances and bin performances:
\begin{equation}
\label{eq:spearman}
    \rho = 1 - \frac{6 \sum_{b=1}^{B} (\mathrm{rank}(\bar{d}_b) - \mathrm{rank}(\pi_b))^2}{B(B^2 - 1)},
\end{equation}
which captures monotonic (not necessarily linear) relationships and is robust to outlier bins. Agreement between $r$ and $\rho$ strengthens confidence in the degradation trend.

Beyond measuring the \emph{presence} of degradation (via correlation sign and significance), we quantify the \emph{rate} of degradation using the linear regression slope:
\begin{equation}
\label{eq:slope}
    \gamma = \frac{\sum_{b=1}^{B} (\bar{d}_b - \bar{d})(\pi_b - \bar{\pi})}{\sum_{b=1}^{B} (\bar{d}_b - \bar{d})^2},
\end{equation}
which represents the expected change in performance per unit increase in \ac{S2DD}. \edited{Equations~\eqref{eq:pearson} and~\eqref{eq:slope} share the same numerator; $r$ is normalised by both standard deviations to produce a dimensionless correlation in $[-1, 1]$, while $\gamma$ is normalised only by the variance of $\bar{d}$ to produce a slope in units of performance per unit distance.} A more negative $\gamma$ indicates faster performance degradation with distance, providing a complementary characterisation of model fragility: two models may have similar correlation magnitudes but different degradation rates, with the steeper slope indicating greater sensitivity to distributional shift.

\subsection*{CaliPPer Performance Prediction}
\label{sec:perf_pred}

To predict aggregate model performance on new, unlabelled datasets from binned distance-performance data, we fit a parametric curve $f: \mathbb{R} \to [0, 1]$ mapping distance to performance from $T$ known evaluation settings (e.g., cross-validation folds or test sets). The following describes the base distance-decay curve; the primary method (density-ratio-corrected prediction below) uses this curve form as a residual correction on top of per-sample density-ratio estimates.

We fit an exponential decay model $f(x) = a e^{-bx} + c$ to the pooled training points $\{(\bar{d}_b^{(t)}, \pi_b^{(t)})\}$ across the $T$ known settings, with $b \geq 0$ enforced to guarantee monotone non-increasing behaviour. Among linear, exponential, and isotonic regression candidates, exponential decay provides a parametric, interpretable model with only three parameters, unlike nonparametric isotonic regression.

\paragraph{Bias-corrected curve (all metrics).}
Different evaluation settings may have different class balances and model calibration characteristics, reflected in the mean predicted probability $\bar{p}$. For threshold-dependent metrics (AP, F1), this vertically shifts achievable performance (AP is bounded below by the positive rate). For ranking metrics (AUROC), although per-sample AUROC is calibration-invariant, subset-level AUROC correlates with $\bar{p}$ through prevalence and cross-test difficulty confounds.

We therefore use a unified bias-corrected curve for all metrics. Let $\bar{p}_b$ denote the mean predicted probability of bin $b$. We fit:
\begin{equation}
\label{eq:vbias_curve}
    f_{\mathrm{bias}}(d, \bar{p}) = a \exp(-b \cdot d) + c + \beta \cdot \bar{p},
\end{equation}
where the bias parameter $\beta$ is fitted jointly with $(a, b, c)$ via L-BFGS-B minimisation of $\mathrm{MSE} + \lambda \beta^2$ (ridge penalty on $\beta$ only; $\lambda = 0.05$). The term $\beta \cdot \bar{p}$ adds a vertical offset that captures calibration variation across datasets. On TCR cross-test, the bias-corrected curve improves AUROC prediction by $+0.19$~R over the plain exponential ($R = 0.75$ vs.\ $0.56$), confirming that the $\bar{p}$ feature is informative even for ranking metrics.

For a new dataset with unknown labels, we bin the query samples by distance and predict the per-bin performance using the fitted curve evaluated at the bin's mean distance with the bin's own mean prediction:
\begin{equation}
\label{eq:pred_perf}
    \hat{\pi}_b = f_{\mathrm{bias}}(\bar{d}_b,\, \bar{p}_b), \qquad
    \hat{\pi} = \frac{\sum_{b=1}^{B} n_b \cdot \hat{\pi}_b}{\sum_{b=1}^{B} n_b},
\end{equation}
where $n_b$ is the number of samples in bin $b$, $\bar{d}_b$ is the mean distance of that bin, and $\bar{p}_b$ is the bin's own mean predicted probability. Each bin's prediction evaluates $f$ at the bin centroid (the same granularity at which the curve was fit) rather than averaging per-sample evaluations $\frac{1}{n_b}\sum_i f(d_i, \bar{p}_b)$, which would introduce a systematic upward bias via Jensen's inequality on the convex exponential term without improving correlation with actual performance. The predicted value $\hat{\pi}$ is clamped to $[0, 1]$.

\paragraph{Density-ratio-corrected prediction.}
To capture within-bin performance variation that the parametric curve cannot represent, we augment the distance-decay prediction with a density-ratio estimation (DRE) corrected per-sample estimate. We train a gradient-boosted density-ratio classifier to distinguish calibration from target features $(d, f)$, then fit a weighted calibrator $c: f \to [0, 1]$ on calibration labels using the resulting importance weights~\cite{bialek2024pape}. For each target sample, $c(f)$ provides a calibrated posterior $\hat{P}(y{=}1 \mid f)$ under the target distribution. The aggregate metric is estimated via the expectation formula:
\begin{equation}
\label{eq:dre_expectation}
    \hat{m} = \frac{1}{n} \sum_{i=1}^{n} \bigl[ c(f_i) \cdot m(\hat{y}_i, 1) + (1 - c(f_i)) \cdot m(\hat{y}_i, 0) \bigr],
\end{equation}
which averages per-sample confusion-matrix contributions weighted by the calibrated posterior $c$. A distance-based residual correction is then fitted on the base estimate residuals (actual metric minus DRE estimate per calibration bin) using dual curve fitting: both exponential decay and right-side Gaussian are fitted to the residuals, and the curve with better $R^2$ is selected with a parsimony threshold ($\Delta > 0.02$) favouring the simpler exponential (4 parameters) over the Gaussian (5 parameters). This distance-based correction captures between-test-set calibration shift that the global density-ratio estimate alone misses, integrating S2DD distance information with the per-sample DRE estimate.

We validate this approach using leave-one-out evaluation: for cross-validation, we hold out one fold, fit $f$ on the remaining folds, and predict the held-out fold's overall performance; for cross-test evaluation, we hold out one test set, fit on the remaining test sets, and predict the held-out set's performance. Prediction accuracy is measured by the mean absolute error between predicted and actual \emph{aggregate} performance values:
\begin{equation}
\label{eq:pred_mae}
    \mathrm{MAE}_{\mathrm{pred}} = \frac{1}{T} \sum_{t=1}^{T} |\hat{\pi}_t - \pi_t|,
\end{equation}
where $\hat{\pi}_t$ is the predicted overall performance for held-out setting $t$ (Eq.~\ref{eq:pred_perf}) and $\pi_t$ is the actual observed aggregate performance on that setting. Note that this MAE quantifies the error in predicting a single scalar per evaluation setting (the model's overall performance on the entire held-out dataset), which is distinct from bin-level curve-shape comparison metrics that measure the similarity of degradation curve profiles across test sets.

\begin{algorithm}[t]
\caption{CaliPPer: Performance Prediction}\label{alg:prediction}
\KwIn{Calibration data $\mathrm{cal\_data} = \{(\mathbf{y}_t, \hat{\mathbf{y}}_t, \mathbf{d}_t)\}_{t=1}^{T}$;
      test predictions $\hat{\mathbf{y}}_{\mathrm{new}}$, test distances $\mathbf{d}_{\mathrm{new}}$;
      metric $\pi$}
\KwOut{Predicted aggregate performance $\hat{\pi}_{\mathrm{new}}$}
\smallskip\hrule\smallskip
\tcp{Step 1: Adaptive binning of calibration data}
$n_{\mathrm{min}} \leftarrow \min(n_{\mathrm{pos}}, n_{\mathrm{neg}})$ from pooled cal labels\;
$B \leftarrow \max(4,\, \min(8,\, \lfloor n_{\mathrm{min}} / 8 \rfloor))$\;
\For{each calibration set $t = 1, \ldots, T$}{
  Partition $\mathcal{Q}_t$ into $B$ equal-sized bins by distance\;
}

\tcp{Step 2: Density-ratio per-sample base estimate}
Train DRE classifier $h^*$: cal features $(d, \hat{y})$ vs.\ test features $(d_{\mathrm{new}}, \hat{y}_{\mathrm{new}})$\;
Importance weights: $\hat{w}_i \propto h^*(x_i)/(1 - h^*(x_i))$ for each cal sample\;
Fit weighted calibrator $c: \hat{y} \to [0,1]$ on cal labels with weights $\hat{w}$\;
$\hat{\pi}_{\mathrm{DRE}} \leftarrow$ Eq.~\eqref{eq:dre_expectation} applied to test predictions via $c$\;

\tcp{Step 3: Dual-curve distance correction on DRE residuals}
\For{each cal bin $b$}{
  $r_b \leftarrow \pi_b^{\mathrm{actual}} - \hat{\pi}_b^{\mathrm{DRE}}$ \tcp*{residual per bin}
}
Fit both exp-decay and right-Gaussian forms of $f(d, \bar{p}) = a\,g(d) + c + \beta\,\bar{p}$ to residuals $\{(\bar{d}_b, \bar{p}_b, r_b)\}$, minimising $\mathrm{MSE} + \lambda\beta^2$ with $\lambda = 0.05$ (ridge applied to $\beta$ only; curve params $(a, b, c)$ unconstrained)\;
Select curve with better $R^2$ (parsimony: prefer exp unless Gaussian improves $R^2$ by $> 0.02$)\;

\tcp{Step 4: Final prediction}
$\hat{\pi}_{\mathrm{new}} \leftarrow \mathrm{clamp}\bigl(\hat{\pi}_{\mathrm{DRE}} + f_{\mathrm{curve}}(\bar{d}_{\mathrm{new}},\, \bar{p}_{\mathrm{new}}),\; 0,\; 1\bigr)$\;
\Return{$\hat{\pi}_{\mathrm{new}}$}
\end{algorithm}

\subsection*{CaliPPer Bayesian Recalibration}
\label{sec:bayesian_recal}

\edited{Standard Platt scaling is a strictly monotone function of the raw score and therefore cannot change AUROC. To enable recalibration that adjusts ranking, not only probability calibration, we derive Platt-scaling parameters analytically from distance-dependent positive predictive value (PPV($d$)) and negative predictive value (NPV($d$)) rather than fitting them by maximum likelihood~\cite{platt1999probabilistic}. We model the log-likelihood ratio as linear in logit space, with a per-bin slope-weighting factor $w(d) \in [w_{\min}, 1]$ that gates the contribution of $(a, b)$ relative to the prior:}
\begin{equation}
\label{eq:bayesian_cal}
\edited{P(Y{=}1 \mid p, d) = \sigma\!\bigl(\operatorname{logit}(\pi) + w(d)\,a(d) + w(d)\,b(d) \cdot \operatorname{logit}(p)\bigr),}
\end{equation}
\edited{where $p$ is the raw model-predicted probability, $\pi$ is the calibration class prevalence, and $(a(d), b(d))$ are the distance-dependent Platt-scaling parameters. The form is equivalent to standard Platt scaling~\cite{platt1999probabilistic} when $w(d){=}1$ and $(a, b)$ are distance-independent, but here $(a(d), b(d))$ are derived analytically from distance-dependent PPV and NPV at each distance bin rather than fitted via maximum likelihood across all data. No Markov chain Monte Carlo sampling is required: the parameters $(a(d), b(d))$ are obtained in closed form from per-bin PPV($d$) and NPV($d$) (Eq.~\ref{eq:b_from_ppv}), and the only prior is the positive-class prevalence $\pi$ (taken as the calibration-set prevalence $\pi_{\mathrm{cal}}$; see adaptive threshold paragraph below). Because $(a(d), b(d))$ vary with distance, two samples with identical raw score $p$ but different S2DD distances $d$ receive different posteriors $P(Y{=}1 \mid p, d)$; this breaks the global rank-preservation of standard Platt scaling and is the mechanism by which \ac{CaliPPer} can change AUROC, not only ECE. Given representative scores $\bar{p}_+$ and $\bar{p}_-$ for the positive and negative prediction bins (estimated as the 25th and 75th percentiles of the calibration set's above- and below-threshold scores, respectively):}
\begin{equation}
\label{eq:b_from_ppv}
\edited{b(d) = \frac{\operatorname{logit}(\mathrm{PPV}(d)) - \operatorname{logit}(1 - \mathrm{NPV}(d))}{\operatorname{logit}(\bar{p}_+) - \operatorname{logit}(\bar{p}_-)}, \qquad a(d) = \operatorname{logit}(\mathrm{PPV}(d)) - \operatorname{logit}(\pi) - b(d) \cdot \operatorname{logit}(\bar{p}_+).}
\end{equation}

\edited{The slope-weighting factor $w(d)$ is the bin-level analogue of Youden's J statistic~\cite{youden1950index}, scaled and floored:}
\begin{equation}
\label{eq:slope_weight}
\edited{w(d) = \max\!\bigl(w_{\min},\; \mathrm{clip}(\mathrm{PPV}(d) + \mathrm{NPV}(d) - 1,\; 0,\; 1)\bigr), \qquad w_{\min} = 0.1.}
\end{equation}

\edited{Substituting Eq.~\ref{eq:b_from_ppv} into Eq.~\ref{eq:bayesian_cal} yields $P(Y{=}1 \mid p, d) = \sigma\bigl((1-w(d))\operatorname{logit}(\pi) + w(d)\operatorname{logit}(\mathrm{PPV}(d)) + w(d)\,b(d)\,(\operatorname{logit}(p) - \operatorname{logit}(\bar{p}_+))\bigr)$, which interpolates in logit space between the prior $\pi$ (when $w{=}0$) and the two-point anchored Platt map $\sigma(\operatorname{logit}(\mathrm{PPV}) + b(\operatorname{logit}(p) - \operatorname{logit}(\bar{p}_+)))$ passing through $(\bar{p}_+, \mathrm{PPV})$ and $(\bar{p}_-, 1{-}\mathrm{NPV})$ (when $w{=}1$). The slope $b(d)$ encodes the model's discrimination ability at a given distance: $b(d) < 0$ if and only if $\mathrm{PPV}(d) + \mathrm{NPV}(d) < 1$, i.e., the model performs worse than random in that bin. When this happens, the raw value $\mathrm{PPV}+\mathrm{NPV}-1$ is negative, the inner clip in Eq.~\ref{eq:slope_weight} pins it to zero, and the outer floor activates so $w(d) = w_{\min}$; the recalibration then shrinks toward the prior $\sigma(\operatorname{logit}(\pi)) = \pi$ rather than amplifying a rank-inverting Platt update. The floor $w_{\min}{=}0.1$ preserves a small residual contribution from $(a, b)$ even in the fully-uninformative regime, providing a smooth rather than difficult fallback. When $w(d){=}1$ (PPV and NPV both near 1), $\operatorname{logit}(\pi)$ cancels exactly and the recalibration recovers the pure two-point anchored Platt form, independent of the prior.}

\paragraph{\edited{PPV/NPV classification threshold.}}
\edited{PPV and NPV are defined relative to a classification threshold $\theta$ used to label predictions as positive ($\hat{y}\geq\theta$) or negative ($\hat{y}<\theta$). We use an adaptive threshold $\theta = \max(2\pi_{\mathrm{cal}}-1,\, \min(2\pi_{\mathrm{cal}},\, 0.5))$, where $\pi_{\mathrm{cal}}$ is the calibration set prevalence. This equals $\theta = 0.5$ for most datasets (prevalence $>0.25$), providing a model-independent split, but adjusts downward for extreme class imbalance to prevent degenerate PPV estimation when few predictions exceed 0.5.}

\paragraph{\edited{PPV/NPV estimation via vbias curves.}}
\edited{We estimate PPV($d$) and NPV($d$) for a new test set using the same vbias exponential decay framework used for performance prediction (Eq.~\ref{eq:vbias_curve}), unifying both pipelines under a single methodology. Designated calibration test sets (v3 and v4 for TCR cross-test; A1-A11 and unseen for BCR cross-test; the calibration half for cross-validation) are each split \emph{independently} into $N_{\mathrm{sub}}$ distance-ordered subsets, where $N_{\mathrm{sub}} = \max(4,\, \min(8,\, \lfloor n_{\mathrm{min}} / 8 \rfloor))$ adapts to the minority-class count to ensure $\geq$8 minority-class samples per bin for stable PPV/NPV estimation. For each subset we compute the mean distance, the bin-level mean prediction $\overline{p}_{\mathrm{bin}}$ (the subset's own mean predicted probability), and the observed PPV and NPV at an adaptive threshold $\theta = \max(2\pi_{\mathrm{cal}}-1,\, \min(2\pi_{\mathrm{cal}},\, 0.5))$, which equals 0.5 for prevalence $>0.25$ and adjusts for extreme class imbalance. We then fit:}
\begin{equation}
\label{eq:ppv_vbias}
\edited{\widehat{\mathrm{PPV}}(d) = a_v \exp(-b_v d) + c_v + \beta \,\overline{p}_{\mathrm{bin}},}
\end{equation}
\edited{with analogous curves for NPV and prevalence. Critically, each calibration test set is binned independently (8 subsets per test set, yielding 16 training tuples from v3+v4), preserving the between-test-set variation in $\overline{p}_{\mathrm{bin}}$ that $\beta$ captures. Pooling calibration test sets before binning destroys this signal and degrades calibration. We fit $(a_v, b_v, c_v, \beta)$ via L-BFGS-B minimisation of $\mathrm{MSE} + \lambda\beta^2$. The recalibration does not assume that far-from-training samples are overconfident; this behaviour is a consequence, not an input. When the fitted $\widehat{\mathrm{PPV}}(d)$ and $\widehat{\mathrm{NPV}}(d)$ decrease with $d$ on the calibration set, the derived slope $b(d)$ from Eq.~\ref{eq:b_from_ppv} shrinks toward zero at large $d$, and far predictions are pulled toward neutral automatically; when the calibration curves are flat in $d$, $b(d)$ is approximately constant and the procedure reduces to standard Platt scaling. Algorithmic behaviour is therefore determined by the measured per-bin PPV/NPV rather than by a prior assumption about over- or under-confidence.}

\paragraph{\edited{Ridge regularisation.}}
\edited{We regularise $\beta$ via an $L_2$ penalty with domain-specific $\lambda$:}
\begin{equation}
\label{eq:ridge_vbias}
\edited{\min_{a_v,\, b_v,\, c_v,\, \beta}\;\; \tfrac{1}{n}\sum_{s}\bigl[y_s - (a_v e^{-b_v d_s} + c_v + \beta \,\overline{p}_s)\bigr]^2 + \lambda\, \beta^2,}
\end{equation}
\edited{with $\lambda{=}0$ (no regularisation on the PPV/NPV curves, as the per-bin estimates are sufficiently stable with the adaptive bin count ensuring $\geq$8 minority-class samples per bin). At prediction time, each held-out test set is binned into $N_{\mathrm{sub}}^{\mathrm{test}}$ distance bins, where $N_{\mathrm{sub}}^{\mathrm{test}}$ applies the same adaptive formula to test labels (capped at $\lfloor |\mathcal{Q}_{\mathrm{new}}| / 4 \rfloor$ to ensure at least 4 samples per bin), and each sample is assigned its bin's mean prediction $\overline{p}_{\mathrm{bin}(i)}$, matching the fit-time feature granularity. Per-sample PPV/NPV are then computed as $\widehat{\mathrm{PPV}}(d_i, \overline{p}_{\mathrm{bin}(i)})$, providing distance-dependent confidence estimates.}

\edited{The fitted curves are evaluated at each validation subset's mean distance to obtain per-subset PPV, NPV and prevalence estimates, which are then substituted into Eq.~\ref{eq:b_from_ppv} to compute $(a, b)$, into Eq.~\ref{eq:slope_weight} to compute the slope weight $w(d)$, and into Eq.~\ref{eq:bayesian_cal} to produce calibrated probabilities.}

\paragraph{\edited{Within-bin monotonicity via slope weighting.}}
\edited{The intended re-ordering operates across distance bins: two samples with the same raw $p$ but different $d$ are mapped to different posteriors via $(a(d), b(d))$, which is what allows AUROC to improve. Within a distance bin, however, the calibration should remain monotone in $p$, because non-monotone within-bin behaviour reflects noise in the per-bin PPV/NPV estimates rather than a real signal. A practical issue arises when PPV and NPV are estimated independently per sample: the derived slope $b_i$ can vary in sign across samples (e.g., $b \in [-1.2, +1.8]$ with mean $+0.67$), destroying even the within-bin ordering of predictions (Spearman correlation between original and calibrated scores drops to 0.08). We address this with the slope weighting $w(d)$ from Eq.~\ref{eq:slope_weight}, attenuating the calibration at distances where the model lacks discrimination ($\mathrm{PPV}+\mathrm{NPV}<1$) and preventing within-bin rank inversion. The floor ensures that even low-discrimination bins receive minimal calibration rather than none, preserving within-bin monotonicity of the calibrated score function while leaving the across-bin re-ordering of $(a(d), b(d))$ intact. The vbias curve's smooth parametric predictions further avoid per-sample sign reversals. The full procedure is summarised in Algorithm~\ref{alg:recalibration}.}

\begin{algorithm}[!ht]
\caption{\edited{CaliPPer: Bayesian Recalibration}}\label{alg:recalibration}
\KwIn{\edited{Calibration data $\mathrm{cal\_data} = \{(\mathbf{y}_t, \hat{\mathbf{y}}_t, \mathbf{d}_t)\}_{t=1}^{T}$;
      test labels $\mathbf{y}_{\mathrm{new}}$, predictions $\hat{\mathbf{y}}_{\mathrm{new}}$, distances $\mathbf{d}_{\mathrm{new}}$;
      optional $\mathrm{train\_anchor}$}}
\KwOut{\edited{Recalibrated probabilities $\tilde{\mathbf{y}}_{\mathrm{new}}$}}
\smallskip\hrule\smallskip
\tcp{\edited{Step 1: Adaptive threshold and bin count}}
\edited{$\pi_{\mathrm{cal}} \leftarrow$ mean of pooled cal labels;\quad
$n_{\mathrm{min}} \leftarrow \min(n_{\mathrm{pos}}, n_{\mathrm{neg}})$ from pooled cal labels}\;
\edited{$\theta \leftarrow \max(2\pi_{\mathrm{cal}}-1,\, \min(2\pi_{\mathrm{cal}},\, 0.5))$}
\tcp*{\edited{adaptive; equals 0.5 for $\pi > 0.25$}}
\edited{$N_{\mathrm{sub}} \leftarrow \max(4,\, \min(8,\, \lfloor n_{\mathrm{min}} / 8 \rfloor))$}\;

\tcp{\edited{Step 2: Collect per-bin PPV, NPV from calibration}}
\edited{$\mathcal{S} \leftarrow \emptyset$}\;
\If{\edited{$\mathrm{train\_anchor}$ provided}}{
  \edited{$\mathcal{S} \leftarrow \{(\mathrm{anchor.d},\, \mathrm{anchor.mp},\, \mathrm{anchor.ppv},\, \mathrm{anchor.npv})\}$}
  \tcp*{\edited{inject training reference}}
}
\For{\edited{each calibration set $t = 1, \ldots, T$}}{
  \edited{Partition $\mathcal{Q}_t$ into $N_{\mathrm{sub}}$ equal-sized bins by distance}\;
  \For{\edited{each bin $s$}}{
    \edited{Compute $\mathrm{PPV}_s$, $\mathrm{NPV}_s$ at threshold $\theta$; $\overline{p}_s \leftarrow \mathrm{mean}(\hat{\mathbf{y}}_s)$}\;
    \edited{$\mathcal{S} \leftarrow \mathcal{S} \cup \{(\bar{d}_s,\, \overline{p}_s,\, \mathrm{PPV}_s,\, \mathrm{NPV}_s)\}$}\;
  }
}

\tcp{\edited{Step 3: Fit vbias curves for PPV and NPV}}
\edited{Fit $\widehat{\mathrm{PPV}}(d, \overline{p}) = a e^{-b d} + c + \beta\,\overline{p}$ to $\mathcal{S}$ via ridge ($\lambda\beta^2$)}
\tcp*{\edited{Eq.~\eqref{eq:ppv_vbias}}}
\edited{Fit analogous curve for $\widehat{\mathrm{NPV}}(d, \overline{p})$}\;
\edited{$\bar{p}_+ \leftarrow Q_{0.25}(\{\hat{y}_i \geq \theta\})$;\quad
$\bar{p}_- \leftarrow Q_{0.75}(\{\hat{y}_i < \theta\})$}\;

\tcp{\edited{Step 4: Apply to test predictions}}
\edited{$\pi \leftarrow \pi_{\mathrm{cal}}$}
\tcp*{\edited{use calibration prevalence (no test labels needed)}}
\edited{$N_{\mathrm{sub}}^{\mathrm{test}} \leftarrow \max\!\bigl(4,\, \min(8,\, \lfloor n_{\mathrm{min}}^{\mathrm{test}} / 8 \rfloor,\, \lfloor |\mathcal{Q}_{\mathrm{new}}| / 4 \rfloor)\bigr)$}
\tcp*{\edited{same adaptive formula on test labels, capped by test size}}
\edited{Partition $\mathcal{Q}_{\mathrm{new}}$ into $N_{\mathrm{sub}}^{\mathrm{test}}$ bins by distance}\;
\For{\edited{each bin $s$}}{
  \edited{$\overline{p}_s \leftarrow \mathrm{mean}(\hat{\mathbf{y}}_s)$;\quad
  evaluate $\widehat{\mathrm{PPV}}_s$, $\widehat{\mathrm{NPV}}_s$ from fitted curves}\;
  \edited{$b_s \leftarrow \frac{\operatorname{logit}(\widehat{\mathrm{PPV}}_s) - \operatorname{logit}(1 - \widehat{\mathrm{NPV}}_s)}{\operatorname{logit}(\bar{p}_+) - \operatorname{logit}(\bar{p}_-)}$;\quad
  $a_s \leftarrow \operatorname{logit}(\widehat{\mathrm{PPV}}_s) - \operatorname{logit}(\pi) - b_s \operatorname{logit}(\bar{p}_+)$}
  \tcp*{\edited{Eq.~\eqref{eq:b_from_ppv}}}
  \edited{$w_s \leftarrow \max\!\bigl(0.1,\, \mathrm{clip}(\widehat{\mathrm{PPV}}_s + \widehat{\mathrm{NPV}}_s - 1,\, 0,\, 1)\bigr)$}
  \tcp*{\edited{slope weighting}}
  \For{\edited{each sample $i$ in bin $s$}}{
    \edited{$\tilde{y}_i \leftarrow \sigma\!\bigl(\operatorname{logit}(\pi) + w_s a_s + w_s b_s \cdot \operatorname{logit}(\hat{y}_i)\bigr)$}
    \tcp*{\edited{Eq.~\eqref{eq:bayesian_cal}}}
  }
}
\Return{\edited{$\tilde{\mathbf{y}}_{\mathrm{new}}$}}
\end{algorithm}

\subsection*{Experimental Setup}
\label{sec:experiments}

\subsubsection*{Datasets}
\label{sec:datasets}

\paragraph{TCR-epitope binding.}
\edited{We use a curated TCR--epitope binding dataset containing 40{,}516 samples spanning 783 unique epitopes. Each sample consists of three chains: the epitope peptide sequence, the CDR3$\alpha$ chain, and the CDR3$\beta$ chain, with a binary binding label. Dataset assembly followed a five-step pipeline (Scripts/data\_preprocess in the code repository). Step~0: 10x Genomics CD8+ T-cell dextramer data from four healthy donors were denoised by subtracting negative-control UMIs, log-normalising, projecting onto 8 PCA components, building a $k$-nearest-neighbour graph with Leiden clustering (hyperparameters selected via random search maximising the normalised mutual information between cluster labels and per-cell max-binder epitope), then filtering to clusters in which $>$92\% of cells bound the same peptide--MHC dextramer (a stringent empirical threshold ensuring that each retained cluster could be confidently assigned to a single epitope), and aggregating to clonotype level. Step~1: the denoised clonotypes were merged with VDJdb~\cite{vdjdb2020} and in-house paired SARS-CoV-2 TCR data, then filtered to paired human $\alpha\beta$ TCRs (requiring both CDR3$\alpha$ and CDR3$\beta$). Antigen specificity in the three training sources above was determined by dextramer staining with UMI denoising (10x Genomics), tetramer-sort flow cytometry (in-house Dong Lab data; six SARS-CoV-2 epitopes), and the heterogeneous pre-curated annotations from VDJdb (a mixture of tetramer, dextramer and other assays); samples were merged without further assay-based filtering, consistent with standard practice for these databases. The McPAS-TCR~\cite{mcpas2017} and IEDB~\cite{vita2019iedb} external test sets (used as held-out cross-tests; see below) were used as-is, applying the same convention of retaining native heterogeneous annotations. Gene nomenclature was standardised and functional TCR sequences validated using pyrepseq. Step~2: to prevent data leakage, TCRs with combined $\alpha$+$\beta$ CDR3 edit distance $<$4, identical V$\alpha$/V$\beta$ genes, and the same epitope specificity were grouped into clusters via connected components; clusters were assigned to five hierarchical folds with a rebalancing step to equalise fold sizes. Step~3: CDR1 and CDR2 sequences were assigned from a V-gene lookup table, and MHC contact pseudo-sequences were derived from IMGT-aligned allele data. Step~4: samples lacking complete CDR or MHC annotations were excluded. CDR3 sequences were conditionally trimmed to remove conserved IMGT flanking residues (N-terminal cysteine and C-terminal phenylalanine/tryptophan) where present. The base split (folds 0--2) contains only positive (binding) samples; negative samples were generated by epitope shuffling (cross-joining CDR features with mismatched epitope and MHC contacts) at a 5:1 negative-to-positive ratio for the validation (fold~3) and test (fold~4) splits only. For evaluation, we then re-split the assembled data into two protocols: (i)~\textbf{5-fold epitope-stratified cross-validation}, in which epitopes are assigned to folds such that each fold contains a distinct set of epitopes, yielding approximately 28{,}000 training and 8{,}000 test samples per fold (test epitopes are entirely unseen during training); and (ii)~\textbf{cross-test evaluation}, in which a single model is trained on three of five cross-validation folds (12{,}066 samples, 476 epitopes) and evaluated on six test sets of increasing divergence: four derived from the assembled dataset (a held-out subset of seen epitopes, unseen epitopes from reserved folds, and two held-out validation splits v3 and v4) and two fully external databases processed independently from the assembled dataset: the McPAS-TCR database~\cite{mcpas2017} (13{,}991 samples, 251 epitopes, 90\% epitopes unseen; 268 samples overlapping with training removed by exact-match deduplication on epitope $+$ CDR3$\alpha$ $+$ CDR3$\beta$ $+$ label) and an IEDB-derived~\cite{vita2019iedb} SARS-CoV-2 test set (9{,}562 samples, 160 epitopes, 71\% unseen; zero sample overlap with training; filtered from 121{,}546 total IEDB samples to exclude non-SARS pathogens, on which the model performs near chance (AUROC 0.50--0.54; Supplementary Table~S16), leaving no distance--performance gradient for S2DD or comparable methods to characterise; Supplementary Table~S6).}

\paragraph{BCR-antigen binding.}
\edited{For BCR evaluation, we collected binding data for SARS-CoV-2 infected and QIV vaccinated cohorts with experimentally determined binding labels (no synthetic negatives). SARS-CoV-2 RBD binding data were curated from the CoV-AbDab database~\citemethods{raybould2021covabdab} following the XBCR-net study~\cite{xbcrnet} (${\sim}$12{,}500 antibody--antigen pairs, 37 RBD variants). Influenza HA binding data were compiled from 17 published studies~\citemethods{joyce2016vaccine,liu2019h7n9,andrews2015immune,henry2019trivalent,dugan2021profiling,whittle2014flow,henry2019influenza,wrammert2011broadly,andrews2017preferential,cheung2020generation,huang2015focused,guthmiller2021broadly,guthmiller2021scitranslmed,bajic2019influenza,kanekiyo2019mosaic,dreyfus2012highly,kallewaard2016structure} (5{,}545 samples, 43 HA antigen variants). All three chains (heavy, light, antigen variant) are used for \ac{S2DD} computation. We construct two evaluation protocols: (i)~\textbf{5-fold antibody-stratified cross-validation}, in which the SARS-CoV-2 and influenza binding data are pooled (${\sim}$18{,}000 samples, 80 antigen variants) and split into five folds by unique antibody identity (heavy chain), ensuring no antibody appears in both training and test for any fold; heavy chains with greater than 0.9 BLASTP similarity are additionally considered seen antibodies. This tests generalisation to unseen antibodies within the same antigen families. (ii)~\textbf{Cross-test evaluation}, in which a single model is trained on folds 0--3 and evaluated on three independent test sets of increasing divergence: an independent set of SARS-CoV-2 variants not present in CoV-AbDab (A1--A11; 281 samples, 4 variants), unseen SARS-CoV-2 variants reserved from the pool (1{,}256 samples, 24 variants), and influenza HA test data (1{,}226 samples, 40 variants); fold~4 serves as the calibration set (3{,}655 samples from both SARS-CoV-2 and influenza).}

\paragraph{Retrospective studies.}
\edited{We additionally applied \ac{CaliPPer} retrospectively to five previously published binding-prediction studies: deepAntigen~\cite{que2025deepantigen} (TCR neoantigen binding), PanPep~\cite{gao2023panpep} (pan-allele TCR meta-learning), XBCR-net~\cite{xbcrnet} (BCR--antigen binding), BigMHC~\cite{albert2024bigmhc} (MHC-I immunogenicity) and AntibioticsAI~\cite{wong2024antibiotics} (small-molecule antibiotic activity). For each study, the original authors' training, validation and test partitions were preserved unchanged. Calibration data were drawn from one of two configurations. In a \emph{genuine retrospective study} (deepAntigen, XBCR-net, BigMHC), a separately published validation cohort served as the calibration set: deepAntigen's zero-shot validation set, XBCR-net's Panel~1 wild-type binders, and BigMHC's im\_val held-out subset. In an \emph{adapted retrospective study} (PanPep, AntibioticsAI), where no separately published validation cohort was available, a distance-sorted half of the test cohort itself served as the calibration set, simulating a two-stage experimental workflow in which a pilot half is labelled first and used to calibrate predictions for a subsequent campaign on the remaining half: PanPep used a zero-shot peptide-identity halfsplit: the zero-shot test cohort (491 unseen peptides, 0\% overlap with training) was partitioned by unique peptide identity into two disjoint halves (245 peptides, $n=832$ calibration; 246 peptides, $n=882$ test), so that no peptide appears in both halves. Both halves remain unseen during training, so calibration and test are evaluated entirely outside PanPep's training distribution. The calibration curves were additionally anchored using majority-test (seen-peptide) statistics as a training-proxy reference point. AntibioticsAI used a compound-level interleaved halfsplit (calibration sampled from odd-ranked indices of the distance-sorted compound list, test from even-ranked indices), with each compound appearing in only one of the three splits (training, calibration and test). The interleaved partition (rather than a top/bottom split) was chosen so that calibration and test sets span the same range of distances to training, mirroring a two-stage screening workflow in which the pilot batch and the deployment batch are drawn from the same compound library under matched screening conditions; the PPV($d$) and NPV($d$) curves fitted on the pilot batch therefore reflect the operating regime of the deployment batch. Per-study sample sizes, calibration sources and split protocols are listed in Supplementary Table~S1. In both configurations, no test-sample labels were used in fitting the calibration curves, and the calibration and test sets are disjoint. The retrospective panel was scoped to models trained once across many receptors and targets; per-receptor protocols such as PRP-TCR~\cite{wang2026prptcr}, in which a separate model is fine-tuned for each individual TCR with that TCR's experimental data, address a different prediction setting and were not included.}

\paragraph{Evaluation convention asymmetry.}
\edited{The evaluation protocols for TCR--epitope and BCR--antigen binding follow the conventions established in their respective literatures. Published TCR--epitope prediction models, including NetTCR~\cite{nettcr}, ATM-TCR~\cite{atmtcr}, ERGO-II~\cite{springer2020prediction} and TCR-BERT~\cite{wu2024tcrbert}, are designed and evaluated on unseen-epitope test splits, where the test peptides do not appear in training. Published BCR--antigen prediction models, including XBCR-net~\cite{xbcrnet}, DeepAAI~\cite{zhang2022deepaai}, MambaAAI~\cite{liu2024mambaaai}, MINT~\cite{ullanat2026learning} and RLEAAI~\cite{wang2024rleaai}, are evaluated on unseen-variant or unseen-antibody splits within the same antigen family. This convention asymmetry reflects fundamental differences in the biology and data availability of the two domains. TCR--epitope binding involves short peptides (8--11 amino acids; predominantly HLA class~I-restricted CD8$^{+}$ T cells in our TCR dataset (${\sim}$91\% class~I, ${\sim}$9\% class~II by unique peptide length), matching the natural composition of the public TCR databases used) interacting with short CDR3 loops (10--20 amino acids), and curated databases (IEDB~\cite{vita2019iedb}, McPAS~\cite{mcpas2017}, VDJdb~\cite{vdjdb2020}) catalogue thousands of distinct epitopes (783 in our TCR dataset), making pan-epitope generalisation a tractable and central research goal. This remains challenging in practice: Lu et al.~\cite{lu2025assessment} reported that 46.4\% of 50 evaluated TCR--epitope models perform at or below chance on unseen epitopes. BCR--antigen binding involves full antigen sequences (hundreds to thousands of amino acids) that differ substantially across pathogen families, with curated databases containing only tens of distinct antigen scaffolds per family (80 antigen variants of 2 pathogen families in our BCR dataset). With such limited scaffold diversity, pan-antigen generalisation is not achievable with current data, and BCR models are necessarily evaluated on within-family variant tasks. Our TCR and BCR evaluation protocols follow these field conventions exactly, and the resulting difference in absolute model performance (TCR median test-set AUROC $\approx$0.55 versus BCR median $\approx$0.75) reflects the difficulty asymmetry inherent to the two tasks rather than any protocol difference of our own.}

\subsubsection*{Models}
\label{sec:models}

We evaluate ten primary prediction models spanning eight architectural families to test the generality of our distance--performance framework: five TCR--epitope binding models (NetTCR~\cite{nettcr}, ATM-TCR~\cite{atmtcr}, BLOSUM-RF~\cite{pham2023epitcr}, ERGO-II~\cite{springer2020prediction} and TCR-BERT~\cite{wu2024tcrbert}) and five BCR--antigen binding models (XBCR-net~\cite{xbcrnet}, DeepAAI~\cite{zhang2022deepaai}, MambaAAI~\cite{liu2024mambaaai}, MINT~\cite{ullanat2026learning} and RLEAAI~\cite{wang2024rleaai}). Together these span CNN, attention, dual-LSTM, Transformer+SVM, random forest, GCN+CNN, state-space (Mamba) and ESM-2-based hybrid architectures; per-model citation, architecture family, framework, input chains and dataset sizes are listed in Supplementary Table~S1. For \ac{S2DD} evaluation, CDR3$\alpha$ is preserved from the original fold data for all TCR models regardless of whether the model uses it as input, ensuring a consistent 3-chain distance computation across all models. All five BCR models use identical antibody-stratified 5-fold CV splits, enabling direct cross-model comparison on the same data partitions.

\paragraph{Retrospective study models.}
\edited{For the five retrospective studies, we used the original authors' published model weights and prediction pipelines without retraining or fine-tuning, spanning four distinct architectural families (Transformer: deepAntigen, BigMHC; meta-learning MAML: PanPep; CNN: XBCR-net; directed message-passing neural network: AntibioticsAI) and four input modalities (TCR--peptide, BCR--antigen, MHC--peptide and small-molecule chemistry). \textbf{deepAntigen}~\cite{que2025deepantigen} is a Transformer-based TCR neoantigen binding model trained on a curated pan-cancer neoantigen corpus; we used the authors' released checkpoint and applied it to both the zero-shot validation set (calibration) and the independent ImmuneCODE cohort (${\sim}$50{,}000 TCR--epitope pairs, test). \textbf{PanPep}~\cite{gao2023panpep} is a pan-allele meta-learning model for TCR--peptide binding that uses model-agnostic meta-learning (MAML) over per-peptide tasks; we used the authors' single released checkpoint in zero-shot inference mode and evaluated on PanPep's own zero-shot test cohort of 491 unseen peptides, with the majority test cohort of 25 seen peptides reused (under the same checkpoint and inference mode) only to provide a training-proxy anchor for the calibration curves. \textbf{XBCR-net}~\cite{xbcrnet} is the same CNN architecture described above for the primary BCR benchmark, but the retrospective evaluation uses the authors' original Panel~1 wild-type binders as calibration and Panel~2 Omicron-era candidate antibodies as test, rather than the antibody-stratified folds used in our primary benchmark. \textbf{BigMHC}~\cite{albert2024bigmhc} is a Transformer-based MHC-I immunogenicity model trained on a large pan-allelic peptide--HLA corpus; we used the BigMHC-IM checkpoint with the im\_val held-out subset (688 samples) as calibration and the MANAFEST cohort (834 samples, independent cross-dataset) as test. \textbf{AntibioticsAI}~\cite{wong2024antibiotics} is a Chemprop directed message-passing neural network for small-molecule antibiotic activity prediction, trained on the authors' Mycobacterium-screening corpus; we used the released checkpoint applied to the 283-compound main test set. Per-study architecture, training corpus and checkpoint sources are listed in Supplementary Table~S1.}

\subsubsection*{Evaluation Protocol}
\label{sec:eval_protocol}

We report three primary performance metrics: area under the receiver operating characteristic curve (AUROC), average precision (AP), and F1 score. For each evaluation setting, we compute Pearson correlation between bin-level distances and performances and report the correlation coefficient $r$ with its $p$-value as indicators of degradation strength.

For performance prediction, we use leave-one-out evaluation across both cross-validation folds and cross-test sets for all five TCR and five BCR models, reporting predicted versus actual performance and absolute prediction error.

\paragraph{Retrospective evaluation protocol.}
\edited{For the five retrospective studies, the \ac{S2DD} framework is applied entirely post-hoc on the original authors' predicted probabilities, with no model retraining, weight updates or fine-tuning. The distance metric is selected per study to match the sequence or molecular modality of the binding domain, since no single metric is optimal across all input types: \textbf{deepAntigen} uses BLOSUM-Smith--Waterman (BLOSUM-SW) over the 8--11-mer neoantigen peptide; \textbf{PanPep} uses a 2-dimensional BLOSUM distance over peptide and CDR3$\beta$, anchored using majority-test (seen-peptide) statistics as a training-proxy reference point; \textbf{XBCR-net} uses a 3-chain Levenshtein distance over heavy chain, light chain and antigen sequence, matching the primary BCR benchmark protocol; \textbf{BigMHC} uses a 2-chain BLOSUM distance over peptide and MHC pseudo-sequence; and \textbf{AntibioticsAI} uses Morgan-fingerprint Tanimoto distance under the LogDist transformation ($k = 0.1$, $b = 0.1$, $K = 50$) over the chemical-fingerprint representation of each compound. In each case the metric was chosen to match the dominant input modality (short peptides $\rightarrow$ substitution-aware peptide distance; multi-chain antibody $\rightarrow$ multi-chain edit distance; chemical structure $\rightarrow$ fingerprint-Tanimoto), consistent with the modular base-distance design of \ac{S2DD} (Supplementary Note~6, ``Distance formulation''). The bin count $B$ and decision threshold $\theta$ follow the same adaptive formulae defined for the primary benchmark (Binned Evaluation and Bayesian Recalibration subsections above), which accommodate the smaller and more imbalanced retrospective calibration cohorts. The two calibration configurations (genuine retrospective and adapted retrospective halfsplit) and the per-study calibration cohorts are described above (Methods, ``Retrospective studies''). We report AUROC and AP as primary metrics for all studies, and additionally report true-discovery rate at top-$k$ (TDR@$k$), defined as the fraction of true positives among the top-$k$ ranked predictions, for studies in which the practical readout is a prioritisation task (deepAntigen, PanPep, BigMHC and AntibioticsAI). TDR@$k$ is plotted as the full sweep over $k \in \{1, 5, 10, 20, 50, 100\}$ in each panel. XBCR-net's 21-candidate Omicron panel is a separate setting in which $k \ll 100$; it is presented as a reranking scatter (Fig.~\ref{fig:retrospective}l), with TDR additionally reported at $k = 8$, the number of known Omicron-era binders and hence the perfect-retrieval cutoff. Per-study calibration sizes, distance metrics, adapted $(B, \theta)$ values and reported metrics are summarised in Supplementary Table~S1.}


\subsection*{Statistical analysis}

\edited{Pearson and Spearman correlation coefficients are reported with two-sided $p$-values from the standard $t$-test on $n-2$ degrees of freedom; 95\% confidence intervals are derived from Fisher's $z$-transformation. For the per-distance-bin $\Delta$AUROC correlations in Fig.~\ref{fig:recalibration}l,m ($n=8$ bins each), the resulting CIs are $[0.69, 0.99]$ for TCR ($r=0.939$) and $[0.19, 0.96]$ for BCR ($r=0.790$). Wilcoxon signed-rank tests for paired before-versus-after recalibration comparisons (Fig.~\ref{fig:recalibration}f,g) are one-sided (alternative: recalibrated metric exceeds raw metric), reflecting the pre-specified directional improvement claim of \ac{CaliPPer}. Rank-biserial effect size is reported as $r = |Z|/\sqrt{n}$, where $Z$ is the standard normal approximation of the Wilcoxon $W$ statistic; exact $p$-values are used for all significance claims. The TCR-unseen group ($n = 15$) reached significance for both metrics (AUROC: $W = 104$, $r = 0.65$, $p = 0.005$; AP: $W = 97$, $r = 0.54$, $p = 0.018$); TCR-seen ($n = 5$) and BCR per-domain comparisons ($n = 5$) did not reach $p < 0.05$. Recalibration effect sizes are reported as per-test-set differences ($\Delta$AUROC, $\Delta$AP) under identical labels.}

\edited{The primary sources of variation examined are model architecture, evaluation fold, test-set identity, biological domain (TCR vs BCR), and base distance metric. Each model $\times$ fold $\times$ test-set combination is treated as one evaluation observation; the nested structure (predictions within test sets within folds within models) is not formally modelled, and pooled summaries (e.g., $n=400$ in Supplementary Table~S19) treat the pooled observations as independent for descriptive reporting. Per-model breakdowns (Supplementary Tables~S14, S20, S21) support inspection at the model level. The pooled-within-domain Wilcoxon test (Fig.~\ref{fig:recalibration}f,g) takes the model--test-set pair as the unit of pairing, treating model identities as the population over which \ac{CaliPPer}'s directional improvement claim generalises.}

\edited{$p$-values are reported uncorrected; for multi-cell summaries such as the $5\times5$ correlation grids in Fig.~\ref{fig:degradation}k,l, the primary inferential claim is the sign-consistency across cells. Exact $p$-values are reported where computable; thresholded values ($p<0.05$, $<0.01$, $<0.001$) are used in tabulated summaries.}

\newpage
\bibliographystylemethods{unsrt}
\IfFileExists{methods.bbl}{\bibliographymethods{ref_methods}}{%
\section*{Methods References}%
}

\section*{Data Availability}
\edited{TCR--epitope binding data were assembled from VDJdb~\cite{vdjdb2020} and 10x Genomics CD8+ T-cell dextramer datasets (UMI-denoised; four healthy donors, ``CD8+ T cells of Healthy Donor 1--4'' series: \url{https://www.10xgenomics.com/datasets/cd-8-plus-t-cells-of-healthy-donor-1-1-standard-3-0-2}, \url{https://www.10xgenomics.com/datasets/cd-8-plus-t-cells-of-healthy-donor-2-1-standard-3-0-2}, \url{https://www.10xgenomics.com/datasets/cd-8-plus-t-cells-of-healthy-donor-3-1-standard-3-0-2}, \url{https://www.10xgenomics.com/datasets/cd-8-plus-t-cells-of-healthy-donor-4-1-standard-3-0-2}), which are publicly available. In-house paired SARS-CoV-2 TCR data used in the assembly are available from the corresponding author upon reasonable request. External cross-test sets were derived from McPAS-TCR~\cite{mcpas2017} and IEDB~\cite{vita2019iedb}. SARS-CoV-2 antibody--antigen binding data were curated from the CoV-AbDab database~\citemethods{raybould2021covabdab} following the XBCR-net study~\cite{xbcrnet}. Influenza HA antibody binding data were compiled from 17 published studies~\citemethods{joyce2016vaccine,liu2019h7n9,andrews2015immune,henry2019trivalent,dugan2021profiling,whittle2014flow,henry2019influenza,wrammert2011broadly,andrews2017preferential,cheung2020generation,huang2015focused,guthmiller2021broadly,guthmiller2021scitranslmed,bajic2019influenza,kanekiyo2019mosaic,dreyfus2012highly,kallewaard2016structure}. Retrospective study data were obtained from their original publications: deepAntigen~\cite{que2025deepantigen}, PanPep~\cite{gao2023panpep}, BigMHC~\cite{albert2024bigmhc} (Mendeley Data: 10.17632/dvmz6pkzvb) and AntibioticsAI~\cite{wong2024antibiotics}. All processed datasets (with in-house data replaced by anonymised sequence hashes where applicable) and evaluation splits are available in the code repository.}

\section*{Code Availability}

\edited{The \ac{CaliPPer} framework, all data preprocessing scripts (Scripts/data\_preprocess), evaluation pipelines and trained model weights are available at \url{https://github.com/jianqingzheng/calipper}.}

\section*{Acknowledgements}
The authors acknowledge computational resources provided by the CAMS Oxford Institute, University of Oxford. This work was supported by the Chinese Academy of Medical Sciences (CAMS) Innovation Fund for Medical Science (CIFMS), China (grant number: 2024-I2M-2-001-1).

\section*{Author contributions statement}
Conceptualization: J.Q.Z. and H.L.; 
Methodology: J.Q.Z. and H.L.; 
Investigation: J.Q.Z., Z.Y. and Y.Z.;
Software: J.Q.Z. and H.L.;
Validation: J.Q.Z., and H.L.;
Formal analysis: J.Q.Z., Z.Y. and S.F.;
Data curation: Z.Y., S. F., E.A., J.Q.Z. and H.L.; 
Visualization: J.Q.Z., H.L., Z.Y. and Y.Z.;
Writing original draft: J.Q.Z. and H.L.; 
Writing - review \& editing: J.Q.Z., H.L., Z.Y., Y.Z., E.A., X.W., X.C., T.D.; 
Resource: T.D., X.C., X.W., H.L.; 
Supervision: T.D. and X.C.

\section*{Competing interests}
The authors declare no competing interests.

\label{EndOfMain}    

\end{document}